\documentclass[aps,twocolumn,floatfix]{revtex4-1}
\usepackage{amsmath,amssymb,graphicx,cases}
\usepackage{epsfig}
\usepackage{textcomp}
\usepackage{epstopdf}
\usepackage{float}
\usepackage{braket}
\usepackage{natbib}
\usepackage{xcolor}
\usepackage[normalem]{ulem}
\usepackage[english]{babel}
\usepackage{enumerate}

\newcommand{\stkout}[1]{\ifmmode\text{\sout{\ensuremath{#1}}}\else\sout{#1}\fi}

\usepackage[colorlinks=true, urlcolor=blue, anchorcolor=blue, citecolor=blue,filecolor=blue,linkcolor=blue,menucolor=blue]{hyperref}

\newif\ifimportant
\importantfalse

\begin{document}
	\title{Extinctions of coupled populations, and rare-event dynamics under \\ non-Gaussian noise}
	
	\author{Tal Agranov}
	\author{Guy Bunin}
	\affiliation{Department of Physics, Technion---Israel Institute of Technology, Haifa 3200003, Israel}
	
	\begin{abstract}
		The survival of natural populations may be greatly affected by environmental conditions that vary in space and time. 
		We look at a population residing in two locations (patches) coupled by migration, in which the local conditions fluctuate in time. We report on two findings. First, we find that unlike rare events in many other systems, here the histories leading to a rare extinction event are not dominated by a single path. We develop the appropriate framework, which turns out to be a hybrid of the standard saddle-point method, and the Donsker-Varadhan formalism which treats rare events of atypical averages over a long time. It provides a detailed description of the statistics of histories leading to the rare event, and the mean time to extinction. The framework applies to rare events in a broad class of systems driven by non-Gaussian noise.
		Secondly, applying this framework to the population-dynamics model, we find a novel phase transition in its extinction behavior. Strikingly, a patch which is a sink (where individuals die more than are born), can nonetheless reduce the probability of extinction, even if it lowers the average population's size and growth rate.
	\end{abstract}
	
	\maketitle
\section{Introduction}
\label{intro}	
	The extinction of populations extended over space is a central question in evolution, ecology and conservation that has been studied extensively both theoretically and empirically \cite{bartlett_stochastic_1960,nisbet_modelling_2003,lande_stochastic_2003,ovaskainen_stochastic_2010,assaf_wkb_2017,leigh_jr_average_1981,metz_what_1983,holt_population_1985,hardin_asymptotic_1988,hardin_comparison_1988,harrison_correlated_1989,hardin_dispersion_1990,lande_risks_1993,jansen_populations_1998,roy_temporal_2005,	abta_amplitude-dependent_2007,matthews_inflationary_2007,schreiber_interactive_2010,khasin_minimizing_2012,khasin_fast_2012,ottino-loffler_population_2020}. Environmental conditions, that vary in space and time, can play an important role in the survival of these populations \cite{holt_population_1985,hardin_asymptotic_1988,hardin_dispersion_1990,hardin_comparison_1988,harrison_correlated_1989,abta_amplitude-dependent_2007,jansen_populations_1998,matthews_inflationary_2007,metz_what_1983,roy_temporal_2005,schreiber_interactive_2010,ottino-loffler_population_2020}.
	
	Here we look at a model of a population residing in two locations (patches), coupled by migration, and experiencing environmental fluctuations, modeled by noisy growth rates \cite{evans_stochastic_2013,hakoyama_extinction_2005,hening_stochastic_2018}. Whereas the extinction of a single isolated population is well understood \cite{lande_risks_1993,leigh_jr_average_1981,assaf_wkb_2017,ovaskainen_stochastic_2010}, much less is known about extinctions in the two-patch system. We present a comprehensive analytical treatment of this long-standing problem, that holds a number of surprises.
	
\begin{figure*}[]
		\begin{tabular}{ll}
			\includegraphics[width=0.32\textwidth, clip=]{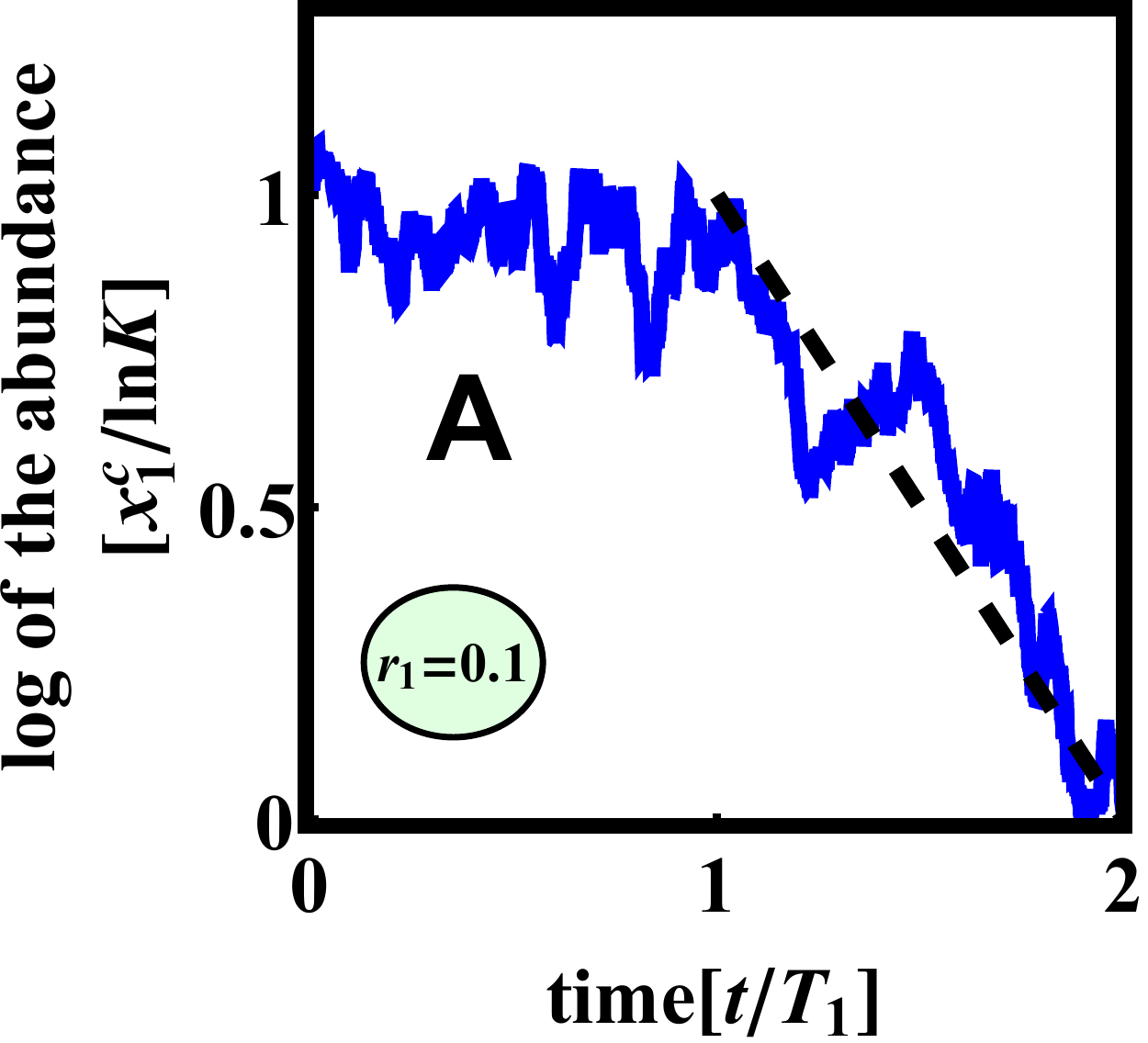}
			\includegraphics[width=0.3\textwidth,clip=]{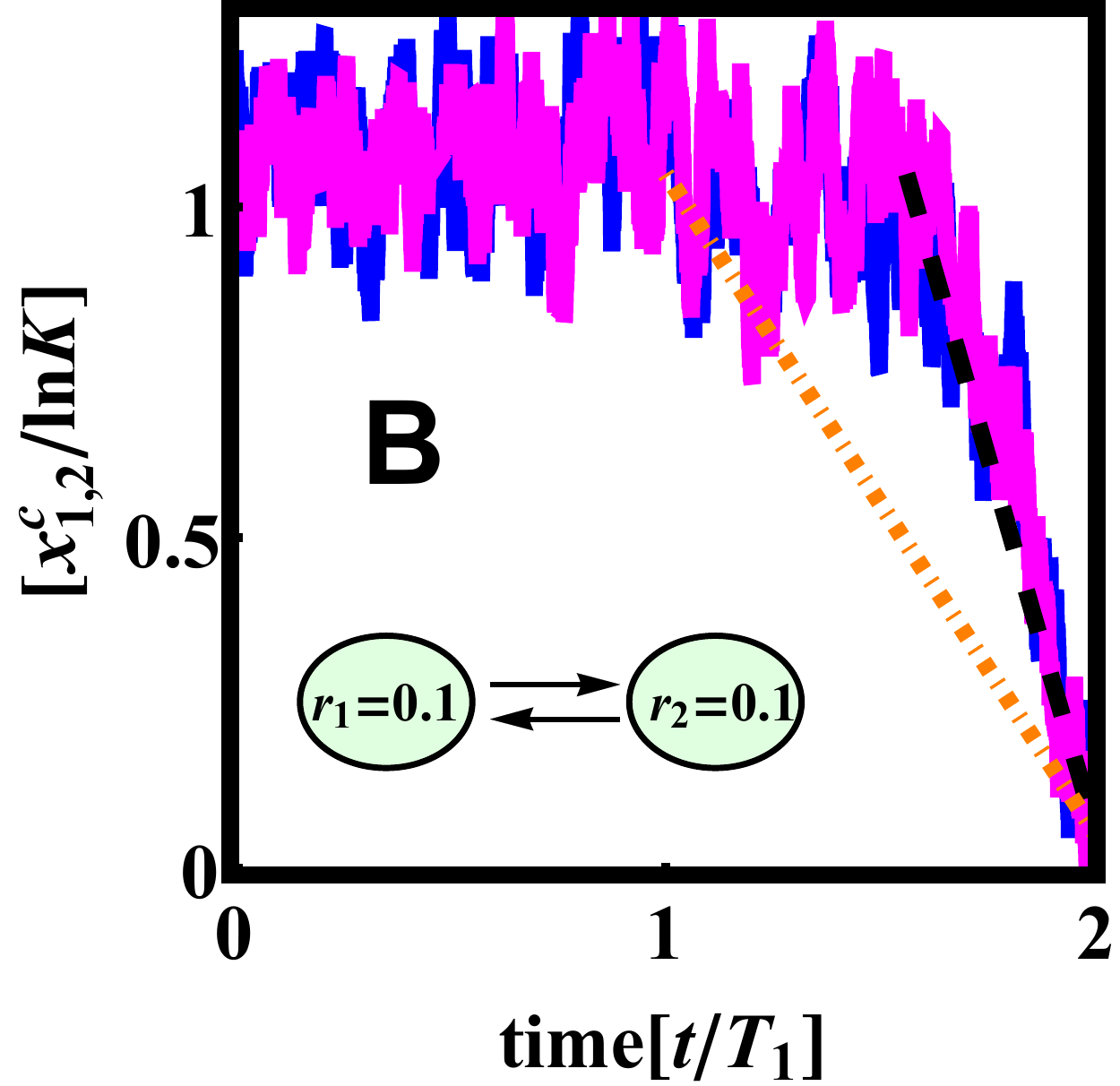}
			\includegraphics[width=0.32\textwidth,clip=]{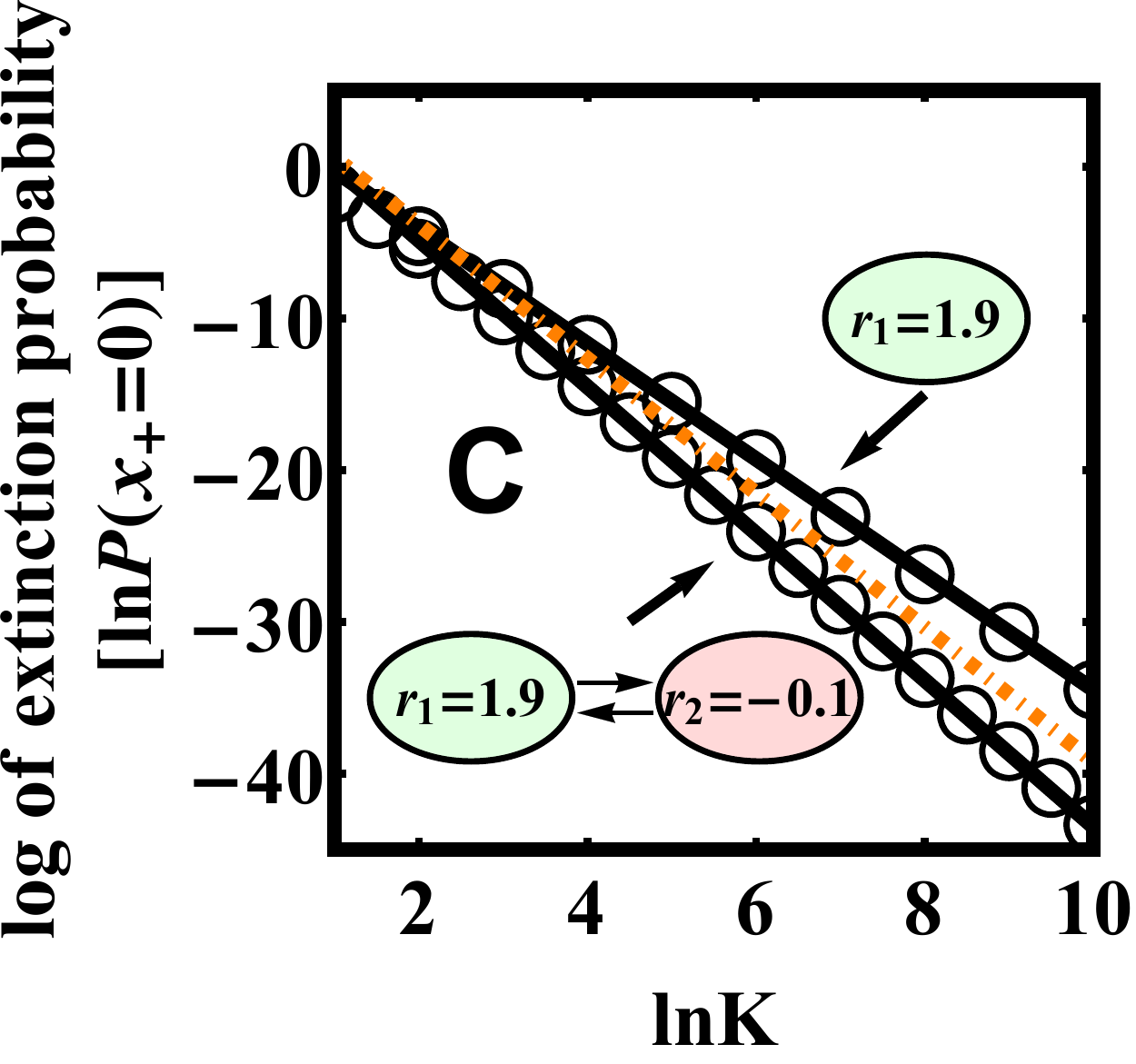}	
		\end{tabular}	
		\caption{ (A) Extinction trajectories and probabilities in a single patch are correctly predicted by the IM. A realization of a system history is shown, that ends with extinction for an isolated single patch. The dashed line is the IM prediction for the decline rate $r_d=r_1=0.1$ where $r_1$ is the growth rate. (B) When two patches are coupled (here $r_2=r_1$), extinction proceeds at a faster decline rate given by our theoretical prediction \eqref{dec} $r_d=0.226$  (black dashed line). The IM prediction for the decline rate which is given by $r_1$ (orange doted line) is incorrect here. In the regime where the population is stabilized by noise-induced stabilization alone ($r_{1,2}<0$) the IM is altogether meaningless, see Sec. \ref{imf}. Other parameters: $T_1=\ln K/r_1$, $\ln K=10$ and $D=0.2$.  (C) Sink habitats can help to protect against extinction: when a source is coupled to a sink, the extinction probability per unit time is \emph{lower} than when it is isolated, see Sec. \ref{ss}. Here extinction  probability for a source patch alone (upper circles), is shown along with a source coupled to a sink (lower circles, indicating lower probabilities). Both are calculated by numerical solutions of the Fokker-Planck equation corresponding to the dynamics \eqref{lan} with a logistic regulating term. Black lines are the analytical predictions \eqref{extinction} for large $\ln K$,
		in perfect agreement with the numerics. The orange doted dashed line is the (incorrect) IM evaluation for the two patch case given by the slope $-2r_1=-3.8$. Here $D=1$.}
		\label{comics}		
	\end{figure*}

	We find a counter-intuitive effect, at small migration rates, where the existence of \textit{sink} patches (where more individuals die than are born) may reduce the probability of extinction, by effectively acting as sources during a potential extinction event, see Fig. \ref{comics}. Thus patches can offer significant protection against extinction even if they have little or detrimental effect on the average population size and growth rates, which are common ecological criteria for survival \cite{howe_demographic_1991,chesson_multispecies_1994,chesson_mechanisms_2000,pande_mean_2020}. The edge of the regime where this happens is marked by a sharp, dynamical phase transition, along with non-analyticity in the large deviation function.
	
	We develop a formalism to derive these and other results. The theory of rare events provides powerful tools to determine the likelihood of extinctions, and how and why they might occur \cite{lande_stochastic_2003,ovaskainen_stochastic_2010,assaf_wkb_2017}. In many systems, rare states such as extinction are reached by a single system history, with negligible probability for all other paths. The formalism used to find this path and its probability is known by various names such as the instanton method (IM), or dissipative WKB \cite{freidlin_random_1998,touchette_large_2009}. 
	
	Yet, we find that extinction events in the two-patch model, as well as rare events in an entire class of other problems that we identify, are \textit{not} reached by a single path as in the IM. To treat this problem, we invoke a different class of rare events, that occur when the long-time average of a given observable attains an atypical value \cite{touchette_large_2009}. Their dynamics are fundamentally different, where a collection of paths are likely, rather than a single one. They are described by the established Donsker-Varadhan (DV) formalism \cite{donsker_asymptotic_1975,donsker_asymptotic_1976,donsker_asymptotic_1983,donsker_asymptotic_2010-1,gartner_large_1977,ellis_large_1984,touchette_large_2009,touchette_introduction_2018}. We find that extinction events can be viewed as a \textit{combination} of the above two classes of rare events, and formulate a hybrid framework that accounts for it, combining the DV and IM formalisms. It allows to evaluate the probability of a rare event, and also to fully characterize the ensemble of system paths which lead to its realization.
	
	The broader class of problems amenable to this formalism includes many systems experiencing colored and in particular non-Gaussian noise, e.g. \cite{klosek-dygas_colored_1989,	hakoyama_extinction_2005,kitada_power-law_2006,kamenev_how_2008,hutt_additive_2008,evans_stochastic_2013,barkai_area_2014,bouchaud_growth-optimal_2015,hening_stochastic_2018,basu_long-time_2019,woillez_activated_2019,	
	walter_first_2020,yahalom_phase_2019,woillez_nonlocal_2020,woillez_active_2020}, sometimes appearing in conjunction with a noise-induced stabilization effect \cite{bouchaud_growth-optimal_2015,evans_stochastic_2013,hakoyama_extinction_2005,hening_stochastic_2018,abta_amplitude-dependent_2007,parker_noise-induced_2011,jansen_populations_1998,parker_noise-induced_2011,valenti_noise_2016,peters_evolutionary_2015}. 
	Within this class, works on specific models provided numerical or partial analytic results \cite{valenti_noise_2016,hakoyama_extinction_2005,kitada_power-law_2006,hutt_additive_2008,abta_amplitude-dependent_2007}, while others \cite{klosek-dygas_colored_1989,kamenev_how_2008,parker_noise-induced_2011,yahalom_phase_2019,woillez_nonlocal_2020,woillez_active_2020} obtained the probability of rare events using the specialized form of certain models.
	We show how to treat the general case and fully characterize the fluctuating dynamics leading to the rare event. 
	
The paper is organized as follows. The coupled population model is presented in Sec. \ref{coupled}. Sec. \ref{single} shows how extinctions in a system comprised of just one patch, can be treaded within the IM. Sec. \ref{imf} shows that the IM fails to treat extinctions in a system of two coupled patches. In Sec. \ref{hyb} we present our theoretical framework which is able to treat extinctions in coupled patches. Secs. \ref{sd} and \ref{lad} present our predictions for the probability and dynamics of extinctions. Sec. \ref{detaildv} provides a review of the DV framework with applications to the extinction populations problem, and Sec. \ref{fpsec} shows an alternative derivation of our hybrid approach based on the Fokker-Planck equation. Lastly, Sec. \ref{conc} presents our conclusions and discusses the broader applicability of our hybrid approach. Details of technical derivations are provided in the Appendices and the SM.
	
	\section{Population residing in coupled patches under environmental noise} \label{coupled}
	Consider two patches harboring populations of sizes $N_{1,2}\left(t\right)$, which grow at rates $r_{1,2}$ at low abundance, and reach a single fixed point at the carrying capacity $K_{1,2}$. $K_{1,2}$ serve as the largest parameter in our problem, and we assume that they both scale with a single large parameter, say $K\equiv\left(K_1+K_2\right)/2\gg1$. The two patches are then coupled by migration. Assuming the dynamics is also subject to white environmental noise, and for large populations where $N_{1,2}\left(t\right)$ can be treated as continuous variables, it is described by the coupled Langevin equations \cite{kamenev_how_2008,ovaskainen_stochastic_2010,assaf_wkb_2017,evans_stochastic_2013,hakoyama_extinction_2005,hening_stochastic_2018} 
	\begin{eqnarray}
	\dot{N}_{1}= & r_{1}N_{1}\left[1-h_1\left(N_1\right)\right]+D\left(N_2-N_1\right)+N_{1}\sigma\eta_{1},\nonumber\\
	\dot{N}_{2}= & r_{2}N_{2}\left[1-h_2\left(N_2\right)\right]+D\left(N_{1}-N_{2}\right)+N_{2}\sigma\eta_{2}.\label{lan}
	\end{eqnarray}
	Here $\!D\!$ is the coupling strength and $\!h\!$ is a regulating term which ensures the growth rates vanish in an isolated patch when $N_i\!\!=\!\!K_i$, such as the logistic term $h\!\!=\!\!\text{sign}\left(r\right)N/K$, but its exact form is irrelevant when addressing extinctions, as their likelihood is dominated by the dynamics away from the fixed point as is shown below. Here $\text{sign}\left(r\right)$ ensures the regulating term suppresses growth above $K$ even when $r$ is negative.
	The right-most terms in Eq. \eqref{lan} model the effect of fluctuating conditions on growth rates (whose average is $r_i$) due to the effect of the environment \cite{lande_stochastic_2003,ovaskainen_stochastic_2010,evans_stochastic_2013}, where $\eta_i(t)$ are zero-mean Gaussian white noises 
	$\left\langle \eta_i\left(t\right)\eta_j\left(t'\right)\right\rangle \!\!=\!\!\delta_{i,j}\delta\left(t-t'\right)$, for $i,j\!\!=\!\!1,2$ \footnote{
		We assume here without lose of generality the Stratonovich convention. In the Ito convention, the values of $r_{1,2}$ in Eq. \eqref{xplos} and on, which enter in all the results below will be different from those in Eq. \eqref{lan}.}. Importantly, the resulting noise terms are multiplicative and scale with population size. 
	 
 	Eq. \eqref{lan} neglects demographic stochasticity. The combined effect of demographic and environmental stochasticity on extinction has been addressed in detail in \cite{kamenev_how_2008} and that it has a sub-leading effect was established there \footnote{The environmental noise term that we consider here corresponds to the regime of \textit{short correlated strong noise} in Ref. \cite{kamenev_how_2008}, that was proven to be dominated by environmental fluctuations alone. See their Eq. (15) and subsequent discussion}. This can 
 	be traced
 	back to the 
 	fact that for large population size demographic stochasticity only scales as $\sqrt{N}$ compared to $\mathcal{O}(N)$ scaling of the environmental fluctuations, see e.g. \cite{lande_risks_1993,kamenev_how_2008,ovaskainen_stochastic_2010}.  Similar reasoning should apply to other sub-leading noise terms, such as additive noise.

	This model and its various extensions has received much attention recently \cite{evans_stochastic_2013,hening_stochastic_2018,hakoyama_extinction_2005}. Much is known about its \emph{typical} behavior--i.e. unconditioned on a rare event like extinction--but far less about extinctions. An equivalent problem appears in economics \cite{bouchaud_growth-optimal_2015}, evolution \cite{peters_evolutionary_2015}, and as a model of a diverse ecosystem, where a species' growth rate fluctuates due to the influence of others \cite{roy_complex_2020-1}.
	
The dynamics \eqref{lan} leads to a stationary distribution $P_s$, peaked around
the carrying capacity $N_i\!=\!K_i\!\gg\!1$. Yet the system can also reach a small number of individuals $N_i\sim\mathcal{O}\left(1\right)$ via a rare noise realization. There the continuous description \eqref{lan} breaks down, and demographic noise may bring the system to extinction, $N_1\!=\!N_2\!=\!0$. 
To leading order, the mean time to extinction (MTE) is given by
$1/P_s\left(N_1\!=\!N_2\!=\!1\right)$ \cite{braumann_growth_2008,kamenev_how_2008,ovaskainen_stochastic_2010}.

\section{Single patch extinction probability given by the IM}\label{single}	
We begin the discussion with uncoupled populations, $D\!=\!0$, where the problem reduces to the classical single patch problem \cite{ovaskainen_stochastic_2010}. Here the MTE has a power law dependence on the carrying capacity, $\text{MTE}\!\sim\! K^{2r/\sigma^2}$ at large $K$, see e.g. \cite{lande_risks_1993,leigh_jr_average_1981}. The exponent $2r/\sigma^{2}$ will be our focus in the following, as it significantly affects the MTE when $K$ is large.

One simple way of arriving at this result is by switching to the logarithmic coordinate $x\!=\!\ln N$, which performs diffusion in a potential
$
\dot{x}\!=\! r\left(1-e^x/K\right)+\sigma\eta
$, here written with the logistic regulating term for concreteness.
The abundance at extinction, where there are $N\!\sim\!\mathcal{O}\left(1\right)$
individuals, corresponds to $x\!=\!0$. Then the MTE is given by the Arrhenius formula for the mean time for $x$ to cross an energy barrier of height  $r\ln K\!\gg\!\sigma^2$ between the metastable fixed point $x\!=\!\ln K$ and extinction at $x\!=\!0$.
	
For what follows, it is instructive to obtain this MTE
using standard IM treatment \cite{kamenev_how_2008,freidlin_random_1998,touchette_large_2009}. Rescaling $y\!=\!x/\ln\! K$ and $\tau\!=\!t/\ln\! K$ gives $
d_{\tau}\!y\!=\! r\!\left(1-K^{y-1}\!\right)\!+\!\sigma\eta/\sqrt{\ln\! K}\simeq r\!+\!\sigma\eta/\sqrt{\ln\! K}
$. The last equality holds during the course of extinction, where $0<y<1$ so $K^{y-1}$ is negligible. 
The IM is valid here due to the small magnitude of the noise $1/\sqrt{\ln\! K}$, and extinction is dominated by a single most probable path  \cite{freidlin_random_1998,touchette_large_2009}. Importantly, the large parameter $\ln\!K$ dose not enter anywhere else in the dynamics.
 This crucial property will be violated in the two patch dynamics, leading to failure of the IM. 

As the Langevin dynamics of a single patch obeys detailed balance, the optimal path is the time-reversal of the typical dynamics, thus following a simple decline at the constant rate $r_d=r$ over the long decline time $T\equiv\ln K /r_d=\ln\!K/r$  \cite{kamenev_how_2008}, see Fig. \ref{comics}(A). Thus for a single patch, the typical growth, the decline rate $r_d$, and the extinction probability are all controlled by the single parameter $r$. As we now show, when coupling two such population patches, a different extinction mechanism comes into play and these three rates differ from one another, as shown in Fig. \ref{comics}(B) and \ref{comics}(C).

To generate the extinction trajectories in Figs. \ref{comics}(A) and (B) we use the time reversed dynamics which correspond to the dynamics \eqref{lan} with logistic regulating term, and initiate it close to the extinction point, see Appendix \ref{fokgen} for details.

\section{The IM fails for non vanishing coupling $D\neq0$}	\label{imf}
Starting from the coupled dynamics \eqref{lan}, it is again helpful to switch to logarithmic coordinates $x_i\!=\!\ln N_i$. We also rescale $D\rightarrow D/\sigma^2$, $r_i\rightarrow r_i/\sigma^2$ and $t\rightarrow t\sigma^2$, resulting in unit noise amplitude $\sigma^2\!=\!1$; the $\sigma$ dependence can always be restored from dimensional considerations. As shown for the single patch (and verified numerically for coupled patches, see Fig. \ref{comics}(B),(C)), extinction is dominated by the dynamics below and not too close to $\ln K$, where the regulating term $h$ is negligible and one obtains \begin{eqnarray}\label{x1}
\dot{x}_1&=&r_1+D(e^{x_2-x_1}-1)+ \eta_1\\
\dot{x}_2&=&r_2+D(e^{x_1-x_2}-1)+ \eta_2\label{x2}.
\end{eqnarray}
Now introduce the sum and difference coordinates 
$
x_{\pm}\!=\!\left(x_{1}\pm x_{2}\right)/2$,
for which
	\begin{eqnarray}\label{xplos}
	\dot{x}_{+} & = & r_{+}+2D\sinh^{2}x_{-}+\eta_{+}/\sqrt{2},\\
	\dot{x}_{-} & = & r_{-}-D\sinh2x_{-}+\eta_{-}/\sqrt{2}.\label{xminos}
	\end{eqnarray}
	Here
	$r_{\pm}\!\equiv\!\left(r_{1}\pm r_{2}\right)/2$,
	and $\eta_{-}$ and $\eta_{+}$ are zero mean and unit variance uncorrelated Gaussian white noises.
	We assume without loss of generality that $r_-\!\geq\!0$.
	Extinction of the population in both patches corresponds to $x_{+}\!=\!0$.
	
	The dynamics of $x_{+}(t)$ are similar to the single-patch case, with growth rate $r_{+}$ and a Gaussian noise of magnitude $1/\sqrt{2}$, but also with an additional fluctuating supplement growth rate 
	\begin{equation}
	0\leq g\left(t\right)\equiv 2D\sinh^{2}x_{-}\label{g}
	\end{equation} 
	that originates from migration between the patches. The
	$x_-$ dynamics, which control this non-Gaussian colored noise, are incorrectly described by the IM during extinction.
	
	This is already evident by looking at the averaged behavior. Here, it is possible for a system to be stable for long times even if the growth rates are negative $r_{1,2}<0$, an effect known as noise-induced stabilization. In such cases the IM fails completely, as we now show.
	
\begin{figure}[]
		\begin{tabular}{ll}
			\includegraphics[width=0.25\textwidth,clip=,trim=0 0 5 0]{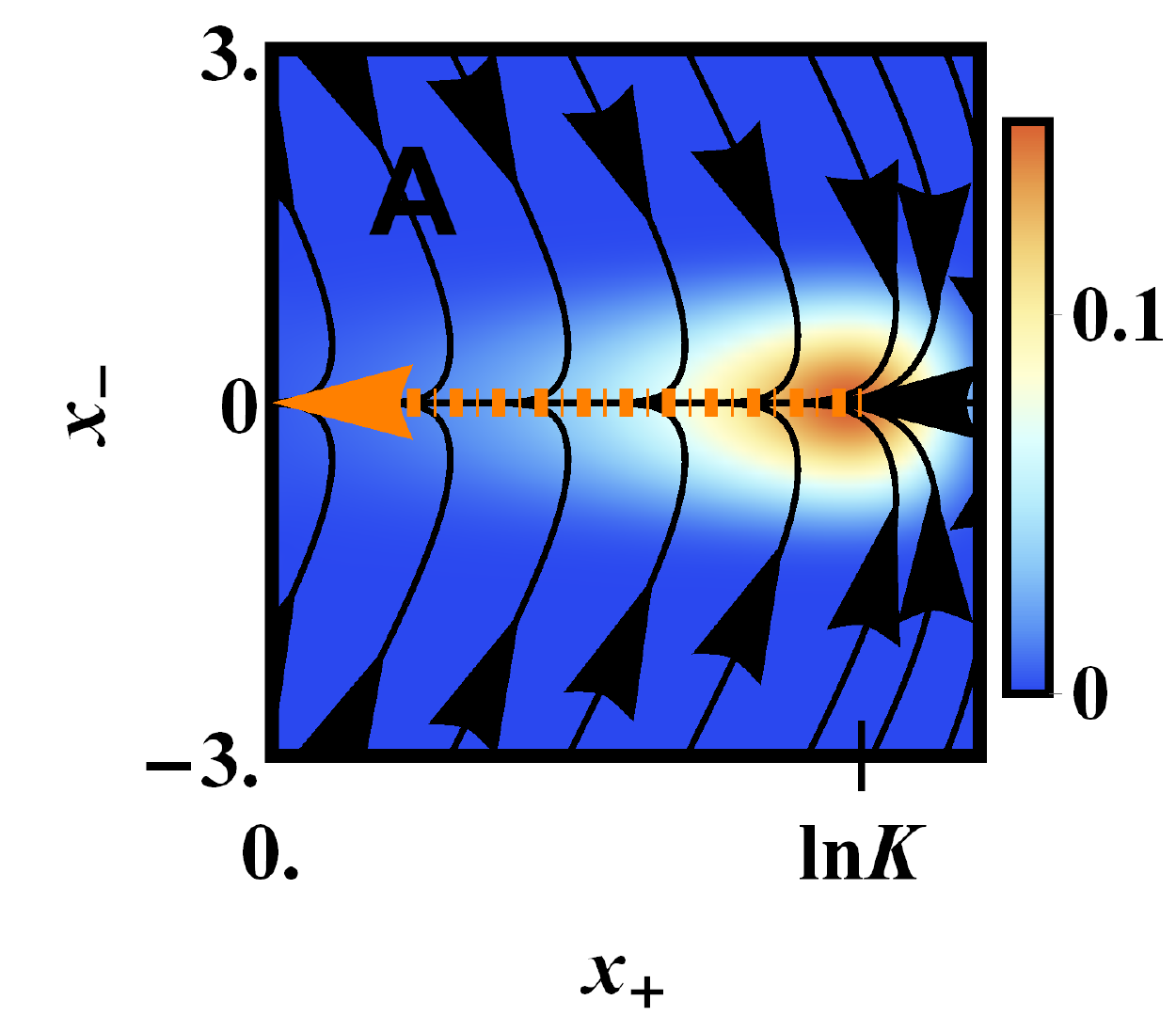}
			\includegraphics[width=0.215\textwidth,clip=]{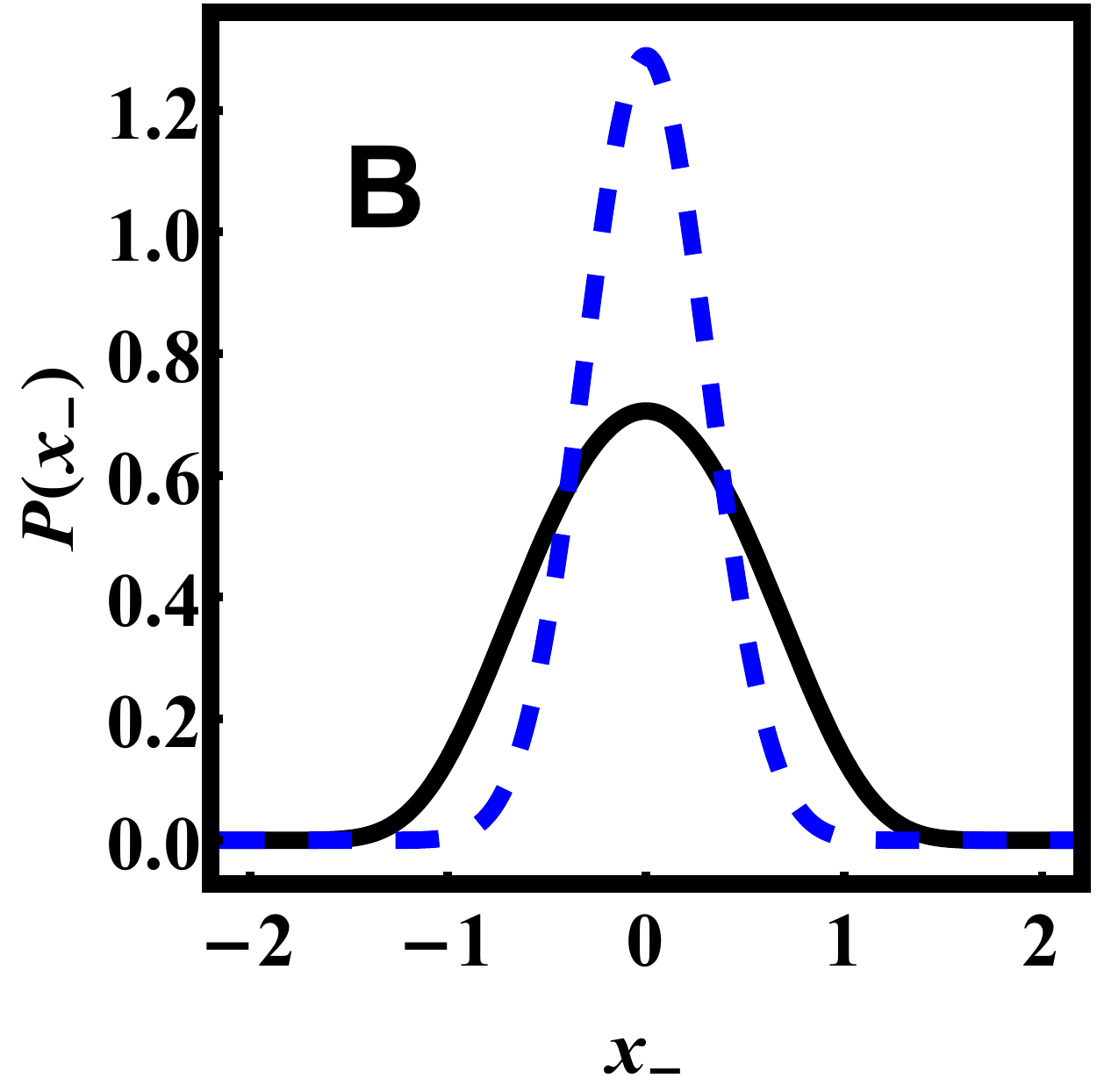}
			
		\end{tabular}	
		\caption{ (A) The stationary joint probability distribution $P_s\left(x_+,x_-\right)$ corresponding to the dynamics \eqref{lan}, with logistic regulating term. Arrows show the direction of the noiseless dynamics. The orange dotted arrow marks the incorrect IM prediction for the extinction trajectory, which simply follows the noiseless dynamics. Even though $r_{1,2}\!<\!0$, the average growth rate is positive, $r_{+}+\left\langle g\right\rangle \!>\!0$, and extinction is a rare process, which proceeds via an ensemble of trajectories.
		Parameters: $r_1\!=\!r_2\!=\!-0.1$, $\ln K\!=\!10$ and $D\!=\!0.3$. (B) The $x_{-}$ distribution during typical growth \eqref{xminosdis} (solid line), is narrower during an extinction event (dashed line). This means that the abundances in the two patches tend to match more closely during an extinction event. Both obtained analytically, at $r_1\!=\!r_2\!=\!2.402$.}
		\label{phase}		
	\end{figure}

	Considering the average total growth rate found by averaging the dynamics \eqref{xplos}, \eqref{xminos} over the $\eta_{\pm}$ fluctuations $
\langle \dot{x}_+\rangle=r_++\langle g\rangle$. 
Whenever $x_+$ starts well-below the carrying capacity, then the difference distribution $P\left(x_-,t\right)$ approaches a stationary distribution $P_-$, during the long growth of $x_+$. This distribution obeys the stationary Fokker-Planck equation $
\partial_{x_-}P_-/4+	\left(D\sinh 2x_- - r_-\right)P_-=0$,
	the solution of which is given by:
	\begin{equation}
P_-\left(x_-\right)=\mathcal N e^{-4\left(D\cosh ^2x_--r_-x_-\right)},\label{xminosdis}
\end{equation}
	with $\mathcal N$ the normalization constant,
	see Fig. \ref{phase}(B). 
	As expected, the distribution of $x_-$ gets narrower when the coupling $D$ is stronger. With this distribution we can now evaluate the supplement growth rate during the long $x_+$ growth 
	\begin{eqnarray}
	\langle g\rangle&=&\int _{-\infty}^{\infty}dx_-2D\sinh^2\left(x_-\right)P_-\left(x_-\right)\nonumber\\	&=&D\left[\frac{\mathcal K\left(1+2r_-,2D\right)}{\mathcal K\left(2r_-,2D\right)}-1\right]-r_-, \label{av}
	\end{eqnarray}
	where $\mathcal K$ is the modified Bessel function of the second kind.
	
	 Now let us consider the case
	when both patches have equal negative growth rates $r_1\!=\!r_2\!\leq\!0$, yet the system can sustain a stable population, which happens whenever the average total growth rate of $x_+$ is positive, $r_++\left<g\right>\!\geq\!0$. 
	This is the celebrated noise-induced stabilizing effect, where fluctuations and migration conspire to stabilize the coupled populations \cite{evans_stochastic_2013,hening_stochastic_2018,jansen_populations_1998,matthews_inflationary_2007}. The mechanism behind this effect is clearly seen in a phase-space portrait of the dynamics \eqref{xplos}-\eqref{xminos} presented in Fig. \ref{phase}(A). 
	
An (incorrect) IM evaluation predicts that the extinction path follows the deterministic dynamics, that is  $x_-\!=\!0$ and $x_+$ declining at rate $-|r_+|$ to extinction. Thus, the IM predicts that the typical dynamics would decline rapidly to extinction, missing out the effect of the noise-induced stabilization which makes extinction rare. 
	One should not expect the IM to apply: a rescaling analogous to that done for a single patch ($y_{\pm}=x_{\pm}/\ln K $ and $\tau=t/\ln K$) will cause $\ln\!K$ to appear not only as a prefactor of the noise. Instead, it also appears in the rescaled dynamics \eqref{xplos}-\eqref{xminos} via the deterministic migration terms, e.g. $2D\sinh^2\left(\ln Ky_-\right)$ in Eq. \eqref{xplos}, which are non-negligible during extinction.

	As it is fluctuations in $x_-$ that stabilize $x_+$ by contributing to the positive supplement growth rate $g$, one expects that during extinction these fluctuations will be suppressed in order to facilitate the decline of $x_+$, see Fig. \ref{phase}(B). This effect is beyond the IM treatment as it involves typical $x_-$ values, yet it can be accounted for by the DV formalism, as is now shown.
\section{Finding extinction probability and dynamics via a hybrid Large deviation approach}	\label{hyb}
\subsection{Conditional extinction for the sum coordinate is given by the IM}
	In the first step of the derivation we find the probability for extinction at a given time $T$, namely to reach $x_+\left(t\!=\!T\right)\!=\!0$ starting from the carrying capacity $x_+\left(t\!=\!0\right)\!=\!\ln K$, \textit{conditioned} on a given $x_-\left(t\right)$ trajectory. With this conditioning, Eq. \eqref{xplos} describes Brownian motion under the \textit{fixed} time-dependent drift $r_++g\left(t\right)$. Extinction at time $T$ corresponds to a generalized Brownian bridge between $x_+\left(t\!=\!0\right)\!=\!\ln K$ and $x_+\left(t\!=\!T\right)\!=\!0$ which in the large $K$ limit is given within the IM formalism. Indeed, starting from the dynamics \eqref{xplos} and employing the rescaling for the sum coordinate alone $y_{+}=x_{+}/\ln K$ and $\tau=t/\ln K$, one arrives at $\dot{y}_{+} = r_{+}+2D\sinh^{2}x_{-}+\eta_{+}\left(\tau\right)/\sqrt{2\ln K}$. Thus an IM treatment for the sum coordinate with a given $x_-(t)$ is valid here,
	and the conditional extinction is dominated by a single optimal path. Finding it and the associated conditional probability is a standard procedure detailed in Appendix \ref{im}. The resulting conditional probability for extinction at time $T$ is given by
	\begin{equation}\label{cond}
	-\ln P\left[x_+\left(T\right)=0|x_{-}\right]\simeq \ln K\left(r_d+G+r_{+}\right)^{2}/r_d,
	\end{equation}
	where we defined the decline rate $r_d=\ln K/T$, and
	\begin{equation}
	G\equiv 2D\int_{0}^{T}\sinh^{2}x_{-}dt'/T,\label{a}
	\end{equation}
	is the time average of the supplement growth rate along the extinction. Note that the probability cost \eqref{cond} only depends on the $x_-$ history via the time average \eqref{a}.  
	\subsection{Total probability for extinction}
\begin{figure}[]
	\begin{tabular}{ll}
		\includegraphics[width=0.21\textwidth,clip=]{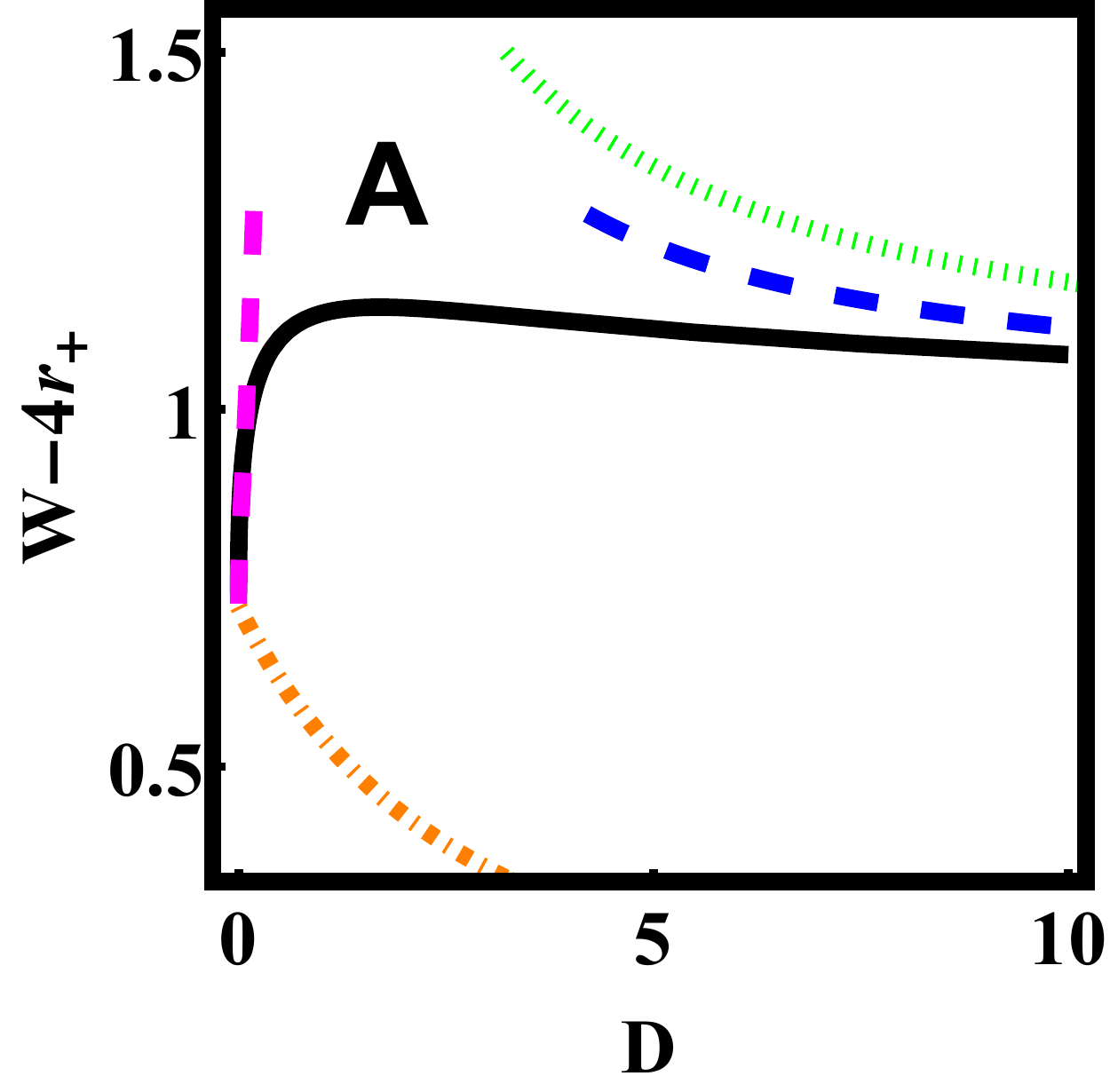}
		\includegraphics[width=0.21\textwidth,clip=]{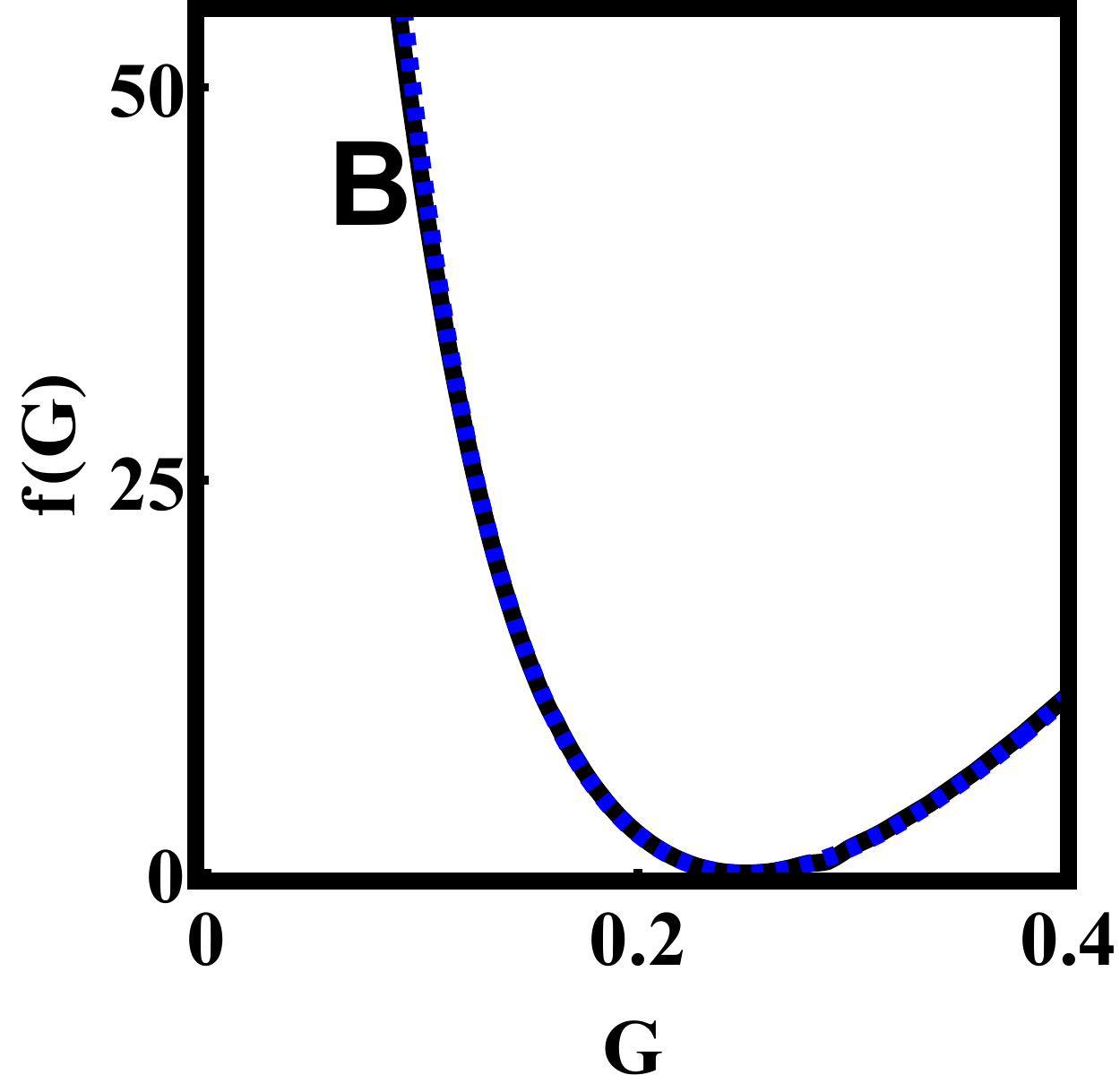}		
	\end{tabular}	
	\caption{(A) The extinction exponent $W$ \eqref{extinction} defined as $\text{MTE}\sim K^W$, as a function of the coupling strength $D$ for $r_1\!=\!2.2$ and $r_2\!=\!0.2$, together with the high \eqref{largedap1} and low \eqref{lowd1} asymptotic expressions in blue and magenta dashed lines respectively. The lower dotted orange line is the incorrect IM prediction obtained by Eqs. \eqref{fim} and \eqref{gim}. The green dotted line is the upper bound in \eqref{bound} given by the expression \eqref{av}. (B) The rate function \eqref{dv} in black solid line for $D=10^2$. The blue dashed line is the high $D$ asymptotics \eqref{ihigh}, showing a perfect match.}
	\label{fh}		
\end{figure}

	We are now in a position to evaluate the total probability for extinction which is \textit{unconditional} on $x_-$ histories. Employing the law
	of total probability, and the fact that the only dependence is via the time average supplement growth rate \eqref{a} we have \footnote{That the conditional probability only depends on $x_-$ via $G$ is in fact exact beyond the IM approximation, and so this equality is exact}
	\begin{eqnarray}\nonumber
	P\left(x_+=0\right)=\int_0^{\infty} dGdTP\left(G,T\right)P\left[x_+\left(T\right)=0|G\right].\\\label{totalprob}
	\end{eqnarray}
	Now we are left with the task of evaluating the probability cost $P\left(G,T\right)$
	of trajectories $\left\{ x_{-}\left(t\right)\right\} $ with a given
	time average value $G$ \eqref{a}. Importantly, as our IM scaling suggest (and is also verified self consistently in the following) the decline time $T$ which dominates extinction scales as $\ln K$, and is thus very long.
Fluctuations of long-time averaged (or empirical) observables is a classic subject in large deviation theory \cite{touchette_large_2009,touchette_introduction_2018}.
Following a large deviation principle, this probability cost decays exponentially with time 
\begin{equation}\label{dv}
-\ln P\left(G,T\right)\simeq Tf\left(G\right).
\end{equation}
$f$ is a convex rate function that attains its minimum at the average value $G\!=\!\left\langle G\right\rangle $, see as an example Figs. \ref{fh}(B) and \ref{eigenplot}(B). The average also coincides with the average of the instantaneous supplement growth rate with respect to the stationary distribution \eqref{av}, $\langle G\rangle\!=\!\langle g\rangle$. One can find $f$ by the established DV formalism \cite{donsker_asymptotic_1975,donsker_asymptotic_1976,donsker_asymptotic_1983,donsker_asymptotic_2010-1,ellis_large_1984,touchette_large_2009,touchette_introduction_2018,gartner_large_1977} reviewed with application to our problem in Sec.\ref{detaildv} and Appendices \ref{dv1} and \ref{wkb}.

Finally, as both the conditional probability \eqref{cond}, and the DV cost \eqref{dv} scale exponentially with $\ln K$, than following the contraction principle \cite{touchette_large_2009}, 
	the integral \eqref{totalprob} is given to leading order
	by 
		\begin{equation}
	-\ln P\left(x_+=0\right)\simeq W\ln K,\label{exprob}
	\end{equation}
	where $W$ is the minimum over $G$ and decline
	rate $r_d$ of the combined probability cost
	\begin{equation}
	W=\min_{G,r_d}\left[\left(r_d+G+r_{+}\right)^{2}+f\left(G\right)\right]/r_d.\label{extinction}
	\end{equation}
	Performing the minimization one finds that
	\begin{equation}
W=-f'\left(G^*\right)\label{GE}
	\end{equation}
with the optimal value $G^*$ found from the solution to	
	\begin{equation}
f'^{2}\left(G^*\right)/4+\left(r_{+}+G^*\right)f'\left(G^*\right)-f\left(G^*\right)=0.\label{opta}
\end{equation}
The corresponding optimal decline rate is given by
	\begin{equation}
	r_d=-\frac{f'\left(G^*\right)}{2}-r_{+}-G^*.\label{dec}
	\end{equation}
		
	Lastly, the MTE is given by the extinction probability \cite{braumann_growth_2008,kamenev_how_2008,ovaskainen_stochastic_2010} and we find that similarly to the single patch, it grows as a power law of the carrying capacity $\langle T \rangle \simeq 1/P\left(x_+=0\right)\sim K^W$. A plot of $W$ appears in Fig. \ref{fh}(A). 
	
	$W$ satisfies a number of simple bounds. An upper bound is given by setting the optimal supplement growth rate to its unconditional value $G=\left\langle G\right\rangle=\left\langle g\right\rangle$ \eqref{av}, where the rate function $f$ vanishes. Plugging it
	in \eqref{extinction} we arrive at the usual
	IM prediction for an isolated single patch but with the modified growth rate $r_{+}\rightarrow r_{+}+\left\langle G\right\rangle$. A lower bound is achieved by neglecting in \eqref{extinction} the cost of $f$ altogether
	and setting $G=0$. This corresponds to the usual IM prediction with growth rate $r_+$, i.e., in the absence of the supplement growth rate. Together, we conclude that 
	\begin{equation}
	0\leq W-4r_+\leq4\left<G\right>.\label{bound}
	\end{equation}
Additional bounds and general properties of the hybrid framework are derived in Appendix \ref{general}.

\subsection{Dynamics conditioned on extinction}
DV provides a recipe to calculate both $f\!\left(G\!\right)$ and the dynamics conditioned on the time average \eqref{a}  as a Langevin process
\begin{equation}
\dot{x}_-^c=-V_G'\left(x_-^c\right)+\eta_-/\sqrt{2},\label{conm}
\end{equation}
where $V_G$ is a modified potential given by the DV theory, see Sec.\ref{detaildv}.
The dynamics of $x_+$ conditioned on extinction, $x_+^c$ can also be derived from mapping to a Brownian bridge. Indeed, \textit{given} a $x_-$ history, the $x_+$ dynamics, conditioned on extinction at time $T$ is a generalized Brownian bridge with a fixed time dependent drift $r_++g\left(t\right)$. In the Appendix \ref{xp} we detail the derivation of its conditioned dynamics that reads
\begin{equation}
\dot{x}_+^{c}=-\frac{x_+^c}{T-t}+\frac{\eta_+}{\sqrt{2}}+\eta_g,\label{conp}
\end{equation} 
where $T\!=\!\ln K/r_d$, and $\eta_g\!=\!2D\sinh^2x_-^c-G^*$ is a zero-mean non-Gaussian noise term that captures the fluctuations in the supplement growth rate $g$ during extinction.
Eqs. \eqref{conm} and \eqref{conp} provide a complete statistical characterization of the trajectories conditioned on extinction. It predicts that extinction is reached by one of a \textit{collection of extinction trajectories}, in which the coordinates $x_{1,2}$ decline together at a rate $r_d$ \eqref{dec}, while their difference fluctuates according to the stationary process \eqref{conm}. 
This is in contrast to the usual IM treatment where only the most probable extinction trajectory is relevant.

We now turn to discuss the predictions of our hybrid formalism to the population dynamics model in the two limits of small and large coupling.

\section{Coupled populations at the small coupling regime $D\rightarrow0^+$}\label{sd}
The small coupling regime displays rich and unexpected behavior.
Consider first the typical behavior where the population grows towards the carrying capacity. As the coupling is small, the two population patches
initially grow as if they were uncoupled and the population difference
$x_{-}$ grows linearly with time. However this cannot proceed indefinitely,
as when $x_-$ grows, the migration term will eventually become non negligible when the difference saturates at a typical value $x_{-}^{*}\simeq\frac{1}{2}\log\frac{2r_{-}}{D}$, at which stage the two patches grow in coordination at the faster rate $r_1$ $(r_1>r_2)$.
Patch 2 contains only a small fraction of the population, and so has negligible effect on the mean and variance of the growth rate, or the total population size. Noise-induced stabilization is negligible, see \cite{evans_stochastic_2013} and the Appendix \ref{tld} for a detailed proof.

Patch 2 might therefore seem to have little bearing on the chances of extinction, as suggested by common ecological criteria \cite{howe_demographic_1991,chesson_multispecies_1994,chesson_mechanisms_2000,pande_mean_2020}, and perhaps even a detrimental effect if it is a sink ($r_2<0$).

Yet we find that even sink patches can significantly reduce the chances of extinction, via a finite ($\mathcal O(D^0)$) effect on $W$. This happens in one of two regimes in $r_1,r_2$, with two qualitatively distinct dynamics preceding extinction that give rise to non-analyticity. 

As we detail in Appendix \ref{ld} and the supplementary material, the DV problem is solved here via a lengthy calculation that invokes matched asymptotics expansion that leads to a surprisingly simple result for the rate function. The rate function $f\left(G\right)$ diverges when $G$ approaches zero over a vanishing boundary layer $G\sim O\left(1/\ln D\right)$, while away from it it is given by the simple parabola
\begin{equation}
f\left(G\right)\simeq \left(G-r_-\right)^2\quad;\quad G>0,\label{lowd}
\end{equation} 
see Fig. \ref{eigenplot}(B).
Using this expression in \eqref{extinction} one finds the optimal supplement growth rate and decline rates:
\begin{numcases}
{\left(G^*,r_d\right)\simeq}
\left(0,\sqrt{\frac{r_1^2+r_2^2}{2}}\right), & $|r_2|\leq r_1$, \label{ginter}\\
\left(|r_+|,r_1\right),& $r_2\leq -r_1$, \label{gstrong}
\end{numcases}
and the corresponding extinction exponent
\begin{numcases}
{W\simeq}
r_1+r_2+\sqrt{2\left(r_1^2+r_2^2\right)}, & $|r_2|\leq r_1$, \label{inter}\\
2r_1,& $r_2\leq -r_1$, \label{strong}
\end{numcases}
see Fig. \ref{smalld1}(B).

Surprisingly, these results agree with a simple IM calculation. To understand these results, and understand this non-trivial coincidence (see also a detailed calculation in Appendix \ref{wkb})), we now turn to describe the extinction dynamics in these two regimes.

\subsection{The extinction sink regime $r_2\!<\!-r_1$}
It is instructive to first look at uncoupled ($D\!=\!0$) patches.
Here extinction only requires conditioning of path 1, which will follow the single-patch instanton decline at rate $-r_1$, as patch 2 declines even without conditioning, and at a faster rate than $-r_1$.

\begin{figure}[]
	\begin{tabular}{ll}
		\includegraphics[width=0.23\textwidth,clip=]{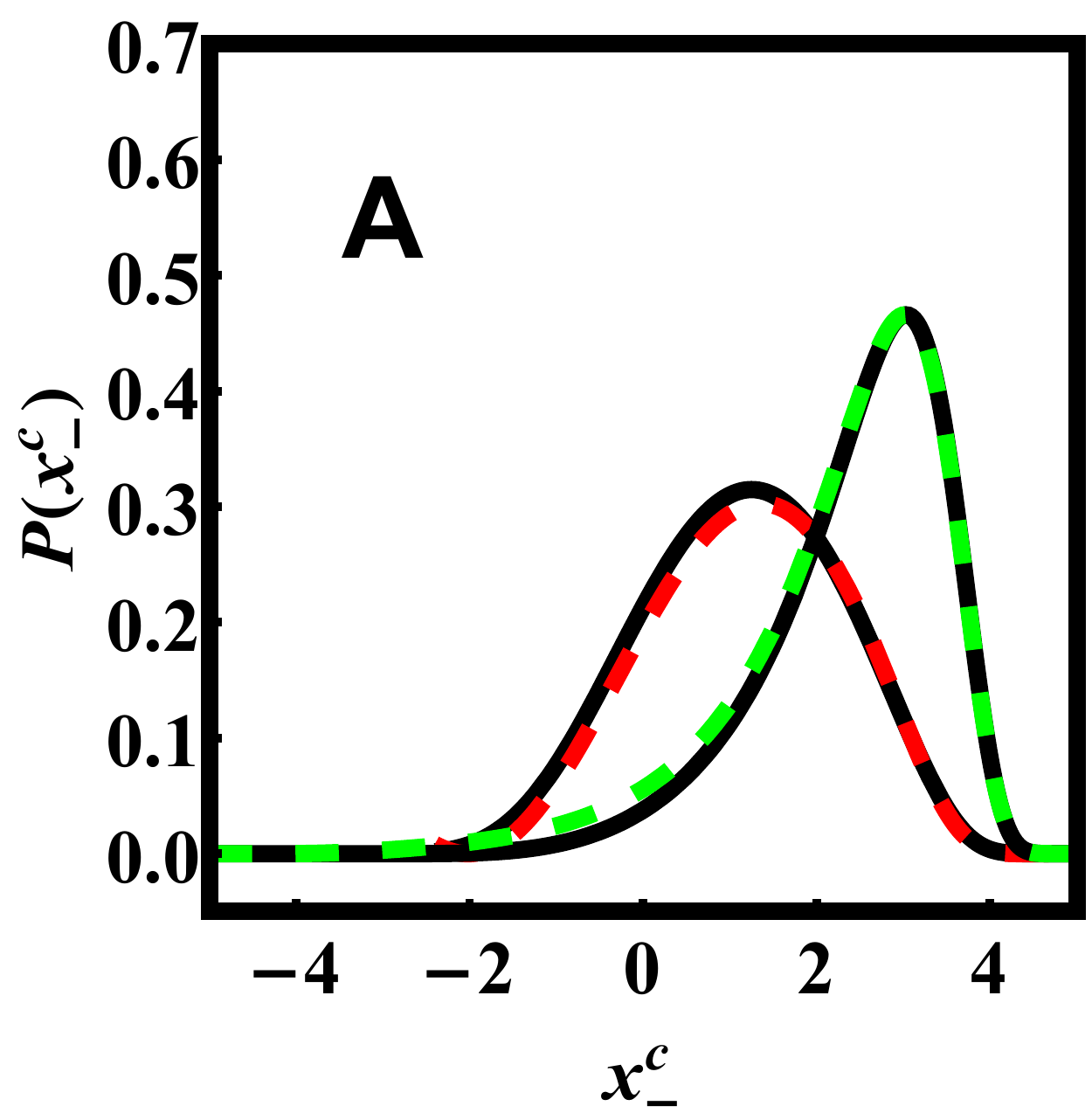}
		\includegraphics[width=0.217\textwidth,clip=]{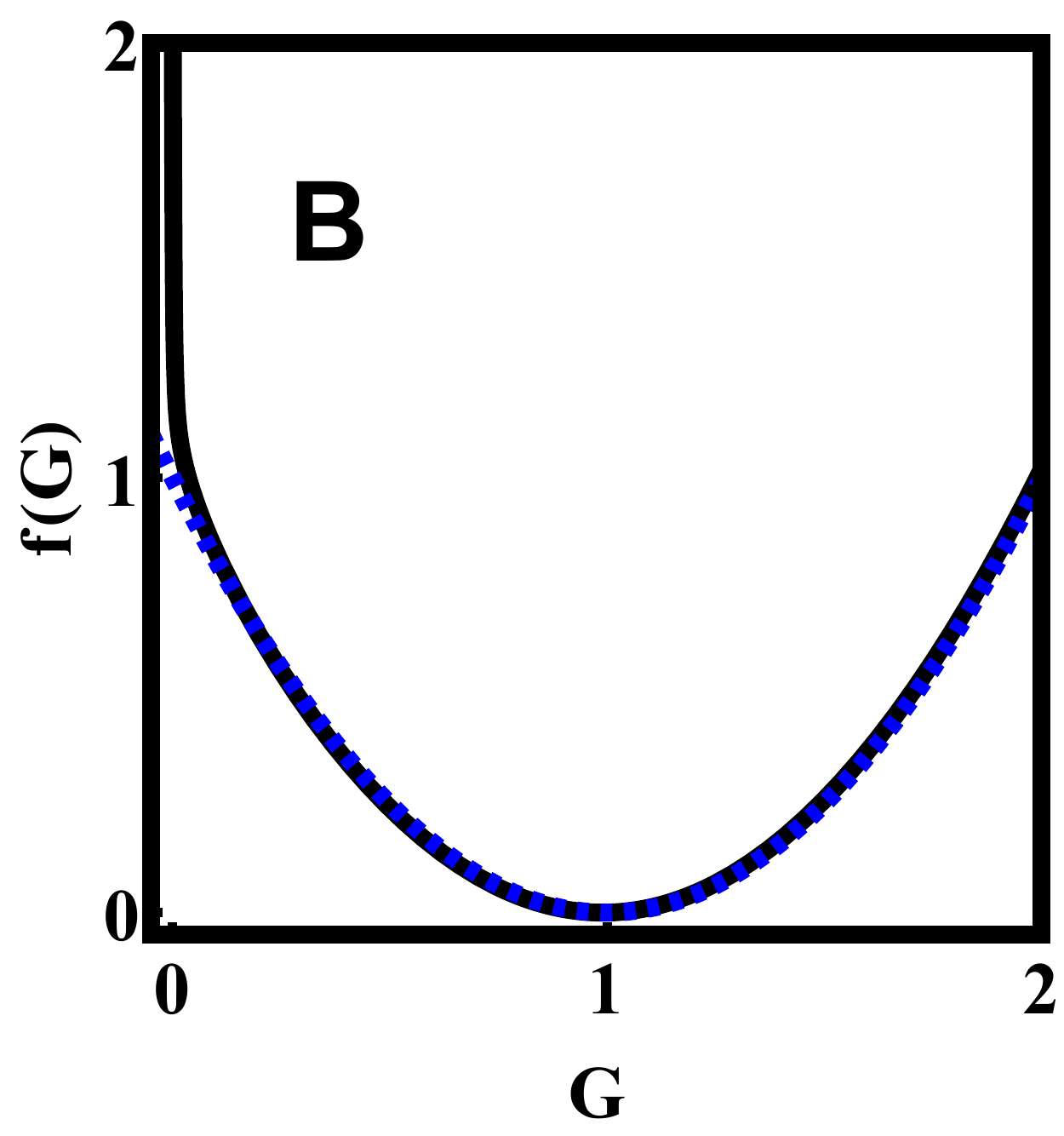}	
		
	\end{tabular}	
	\caption{(A)  the stationary distribution $P\left(x_-^c\right)$ of $x_-^c$, the $x_-$ coordinate conditioned on extinction in black solid lines. The dashed green, and dashed red are the analytical predictions at small $D$, Eqs. \eqref{negr} and \eqref{eigen2} respectively, in the extinction sink regime, with $r_1=4,r_2=-4.41$ (Dashed red), and extinction source regime, $r_1=4,r_2=-3.47$ (Dashed green). (B) The rate function \eqref{dv} in black solid line with $r_-=1$. The blue dashed line is the low $D$ asymptotics \eqref{lowd}. In both panels $D=10^{-2}$. }
	\label{eigenplot}		
\end{figure}

One expects that under small $D\!>\!0$, extinction will still only require the conditioning of patch 1 alone. Indeed, the typical dynamics of patch 2 coupled to the instanton dynamics of patch 1 will reach extinction in coordination with patch 1 at the slower rate $-r_1$, aided by migration. Under such extinction dynamics, patch 2 harbors a much smaller population than in patch $1$, $x_-^c\!\sim\! \ln\left(|r_+|/D\right)$ (Fig. \ref{smalld}(A)), and so patch 1 proceeds with little effect of migration from the much smaller population in patch 2, making this simple picture self consistent. 
This picture is in fact exact, as is shown by taking the limit $D\rightarrow0^+$ on the dynamics towards extinction \eqref{conm}-\eqref{conp}. A rather lengthy calculation, detailed in Appendix \ref{p} gives
\begin{eqnarray}
\dot{x}_1^c&=&-\frac{x_1^c}{T-t}+\eta_1,\label{xc12}\\
\dot{x}_2^c&=&r_2+D\left(e^{x_1^c-x_2^c}-1\right)+\eta_2\label{xc22},
\end{eqnarray}
with $T\!=\!\ln K/r_1$.
That is, $x_1^c$ heads to extinction at rate $r_d\!=\!r_1$ as if migration is absent (a Brownian bridge), while the $x_2^c$ dynamics are unconditioned, except for the migration from patch 1, with given $x_1^c$ trajectories.

\begin{figure}[]
	\begin{tabular}{ll}
		\includegraphics[width=0.2\textwidth]{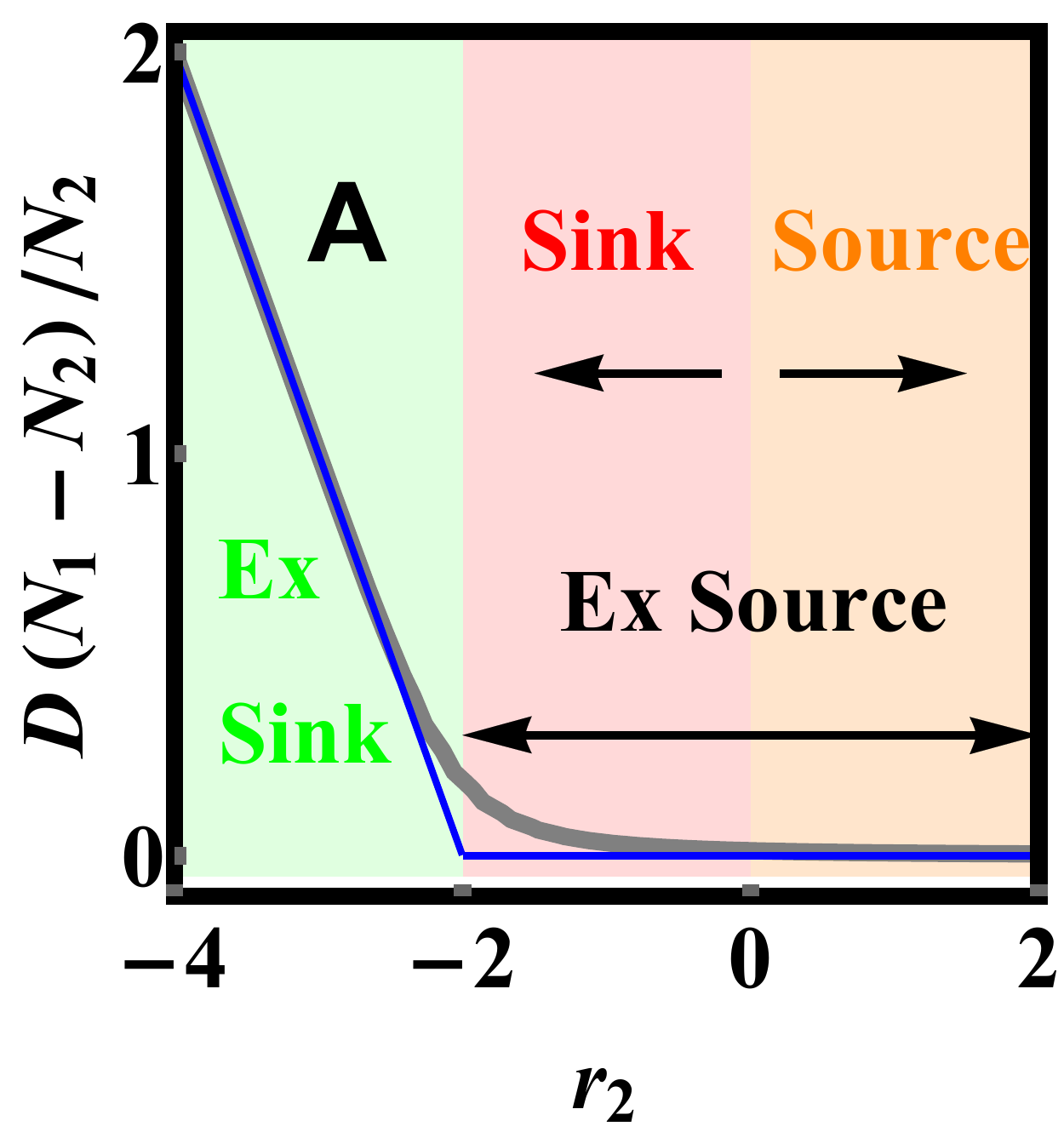}
		\includegraphics[width=0.19\textwidth]{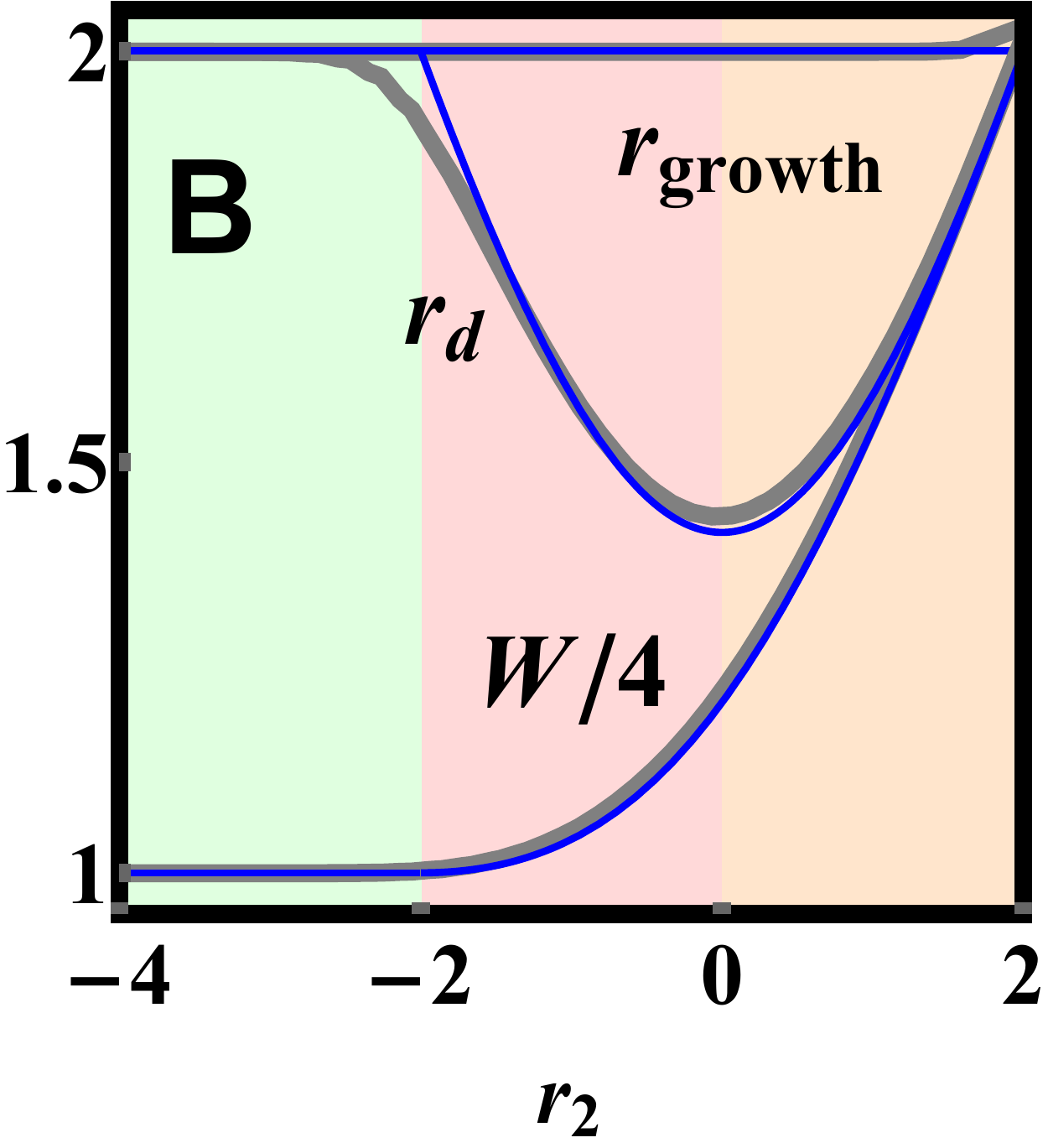}		
	\end{tabular}
	\caption{Extinction at small coupling $D$, showing three regimes. (A) Migration into patch 2 during extinction, normalized by population size, for $r_1\!=\!2$. (B) The extinction probability exponent $W$, decline rate $r_d$ and growth rate. Note that in the extinction sink regime $r_2\!<\!-r_1$, patch 2 acts as a sink also during extinction, in that migration is non-negligible, and $W\!=\!4$ as for patch 1 alone; while in the extinction source regime $-r_1\!<\!r_2$, migration into patch 2 vanishes and the extinction probability is suppressed $W\!>\!4$. This happens both when patch 2 is a sink, $r_2\!<\!0$ (light red), or a source (light orange). The thin blue lines are analytical predictions for $D\rightarrow0^+$, and the thicker gray lines for $D\!=\!10^{-3}$.}
	\label{smalld1}	
\end{figure}

Here patch 2 acts as
a sink much like it would in normal conditions and is supported by migration. Indeed, migration into patch 2 normalized by population size, reaches a finite value $G^*\simeq\langle D(N_1-N_2)/N_2\rangle_c\!\simeq\!|r_+|$
, see Fig. \ref{smalld1}(A) (here $\langle..\rangle_c$ denotes an average with respect to the conditioned dynamics \eqref{conm}). 
The corresponding difference distribution is given here by
\begin{equation}
P\left(x_-^c\right)\simeq\mathcal N e^{-4\left(D\cosh ^2x_--G^*x_-\right)},\label{negr}
\end{equation}
which is simply the unconditioned distribution \eqref{xminosdis} with the shift $r_-\rightarrow G^*\simeq|r_+|$, see Fig. \ref{eigenplot}(A).

To sum it up, in the extinction sink regime, the extinction of the total population is governed by the extinction of the uncoupled patch 1 with $W\simeq 2r_1$, which is why an IM evaluation here turns out to be exact .

\subsection{The extinction source regime, $-r_1\!<\!r_2\!<\!r_1$}\label{ss}

Here, the extinction of an uncoupled patch 1 would proceed at the rate $-r_1$, which is more negative than the typical growth rate in patch 2 alone,  $\left<\dot{x}_2\right>\!=\!r_2\!\geq\! -r_1$. 
So, when coupled, extinction also requires unfavorable conditions in patch 2, \textit{lowering} the chance of extinction. Strikingly, this includes a regime where $r_2$ is negative $-r_1\!<\!r_2\!<\!0$, and during normal conditions acts as a \emph{sink}, see Fig. \ref{comics}(C).  When extinction does
proceed, we find that it happens at an intermediate rate between $r_2$ and $r_1$ given by their root-mean-square
$
r_{d}=\sqrt{\left(r_{1}^{2}+r_{2}^{2}\right)/2}.
$
In addition, we find that the optimal supplement growth rate is vanishingly small $G^*\rightarrow0$ so that migration is negligible during the extinction
event. 
In that sense, patch 2 is no longer a "sink" whose population is supported by migration. On the contrary, it acts against extinction and can be thought of
as a source. Without migration, extinction reduces to the simultaneous decline of two \textit{isolated} patches. Thus, an IM evaluation of the extinction probability is correct here under the additional constraint that the two patches decline simultaneously to zero. 

\begin{figure}[]
	\begin{tabular}{ll}			
		\includegraphics[width=0.23\textwidth]{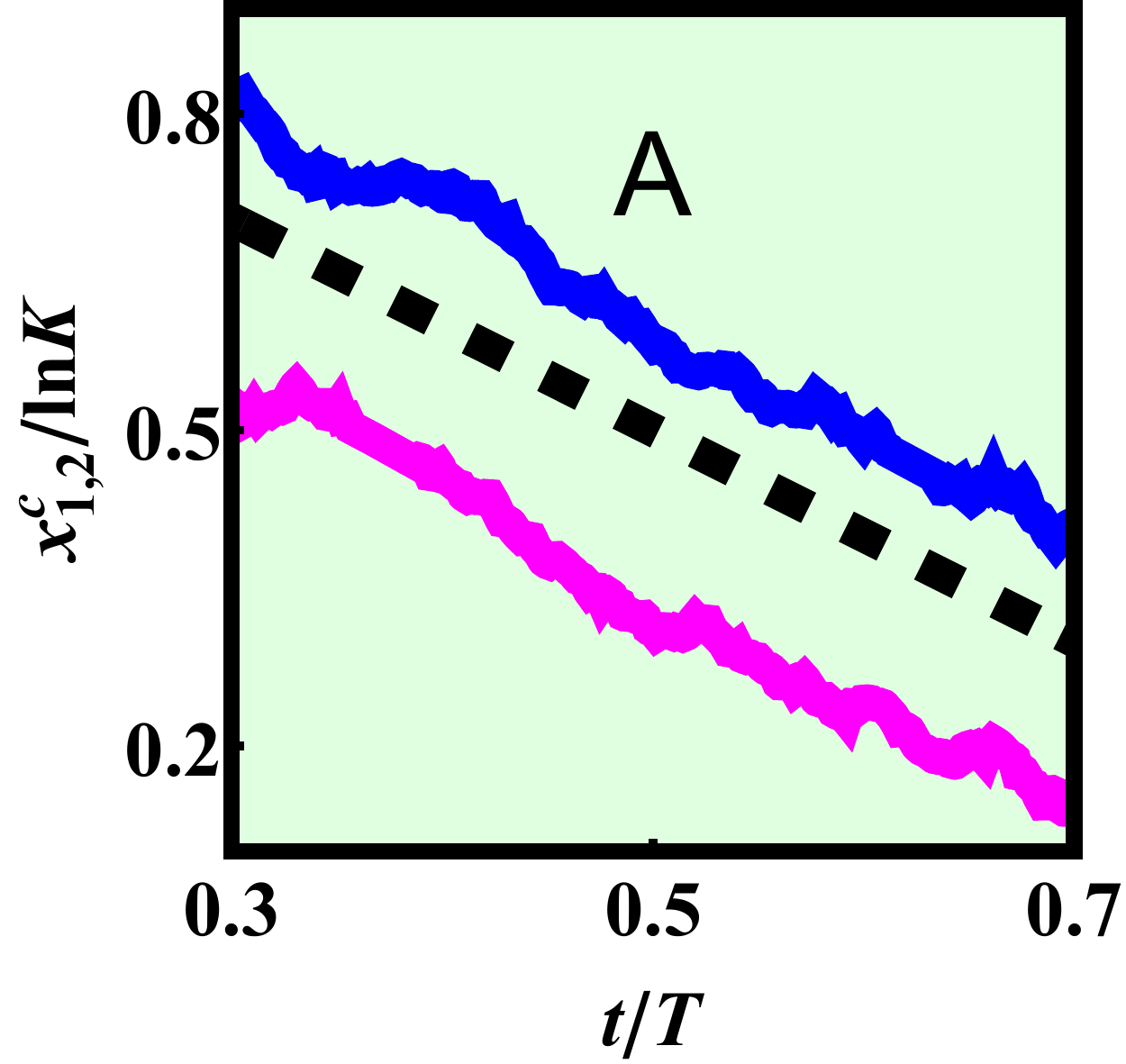}
		\includegraphics[width=0.23\textwidth]{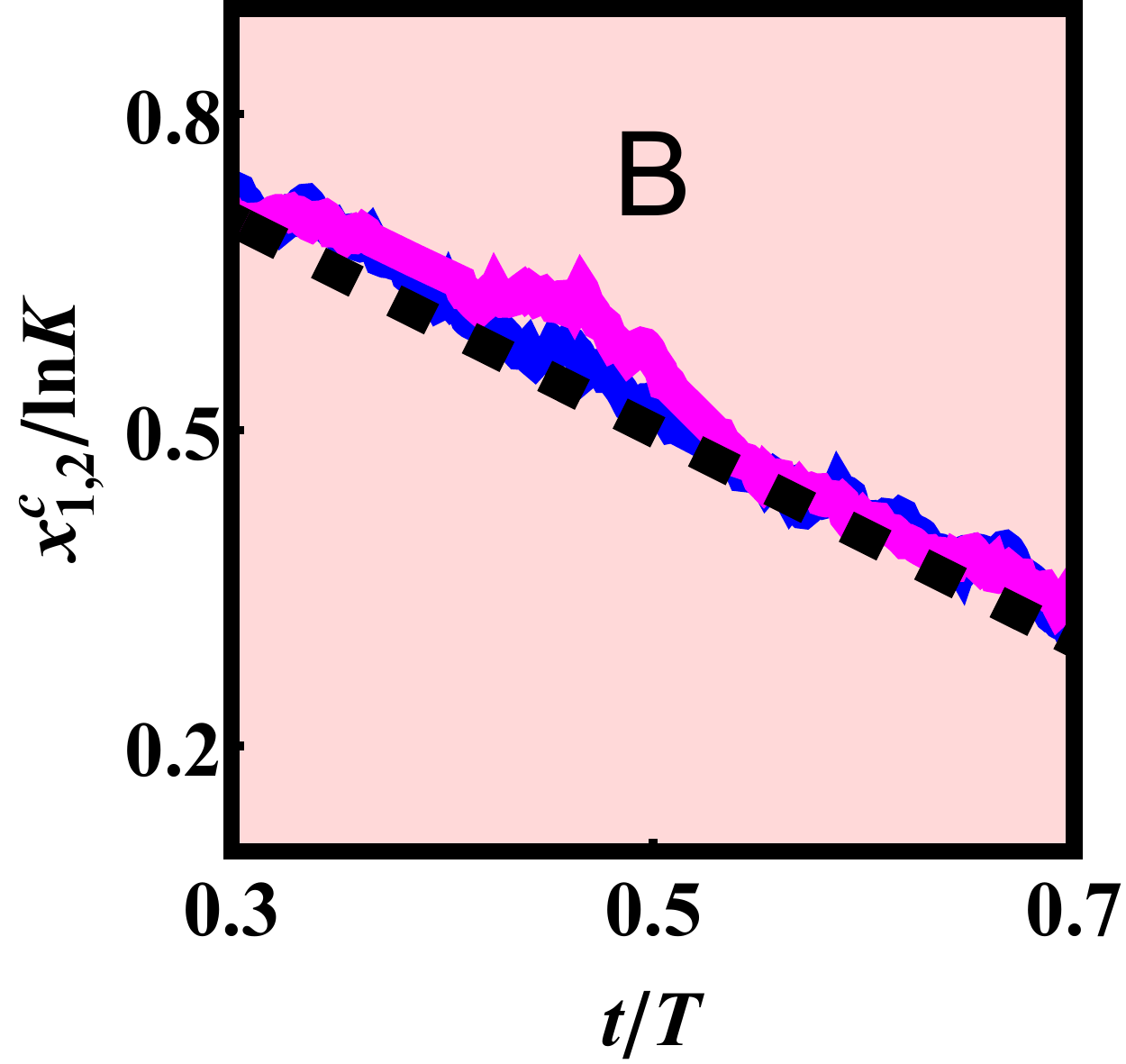}
	\end{tabular}
	\caption{ A realization of extinction trajectories for $x_1^c$ in blue and $x_2^c$ in magenta generated from the process \eqref{conm}-\eqref{conp}. (A) The extinction sink regime ($r_1\!=\!4,r_2\!=\!-4.41$). (B) extinction source regime, for a sink patch 2, $r_2=-3.47<0$. Dashed lines show the predicted decline rate \eqref{dec}. In both $D\!=\!10^{-3},\ln K\!=\!20$.}
	\label{smalld}	
\end{figure}

The conditional  distribution for $x_-^c$ in this regime, derived in Appendix \ref{ld}, is centered near the origin and spreads over a large scale $\sim |\log D|$, see Fig.  \ref{eigenplot}(A). Nevertheless, the corresponding supplement growth rate, evaluated by averaging $g$ \eqref{g} with respect to this distribution, vanishes to leading order $G^*=\mathcal O \left(1/|\log D|\right)$. An example of extinction paths in this regime are shown in Fig.\ref{smalld}(B).

To sum it up, an IM evaluation of $W$ turns out to coincidentally reproduce the correct result \eqref{inter}-\eqref{strong} at small $D$ but due to two different reasons: in the extinction sink regime \eqref{strong} only patch 1 contributes, acting like a single-patch decline. In the extinction source regime \eqref{inter} migration is negligible, and extinction corresponds to the simultaneous decline of two single patches, correctly captured by IM, under the additional constraint that the two patches decline together to zero.

\section{Coupled populations at large coupling - effective single patch}\label{lad}
At strong coupling $D\rightarrow \infty$, the two patches are infinitely coordinated, and the difference coordinate approaches zero $x_{-}\rightarrow0$. Nevertheless, due to the diverging $D$ pre-factor in the supplement growth rate \eqref{g}, the average \eqref{av} approaches a finite value in this limit $\langle G\rangle \simeq1/4$.  A further suppression of the supplement growth rate during extinction necessitates additional extreme suppression of the $x_-$ fluctuations which becomes improbable in this limit, and so the optimal supplement growth rate for extinction must approach its unconditioned average $G^*=\langle G\rangle$. Thus, the upper bound for the extinction
probability \eqref{bound} is saturated  $W\!=\!4\left(r_+ +\langle G \rangle\right)+\mathcal O\left(D^{-1}\right)$. 

This argument is made exact within the explicit solution of the DV problem which we present in Appendix \ref{hd} with account of sub-leading corrections at finite $D$. As we detail in Appendix \ref{hd}, these corrections help identify a regime where the extinction exponent displays a non monotonic dependence on $D$ (see Fig. \ref{fh}(A)). Thus, there is an optimal coupling strength for protection against extinction.

We now return to detail the DV problem which underlays our calculations.

	\section{Details of the The DV problem used in the hybrid framework}\label{detaildv}
	The established DV formalism \cite{donsker_asymptotic_1975,donsker_asymptotic_1976,donsker_asymptotic_1983,donsker_asymptotic_2010-1,ellis_large_1984,touchette_large_2009,touchette_introduction_2018,gartner_large_1977} provides a recipe to calculate both $f\!\left(G\!\right)$ and the dynamics conditioned on the time average \eqref{conm}.
	Finding these is reduced to an effective eigenvalue problem. $f(G)$ is related by a Legendre-Fenchel transform \cite{gartner_large_1977,ellis_large_1984}
	\begin{equation}\label{GE1}
	f\left(G\right) = \sup_k \left[ k G - \xi\left(k\right) \right]
	\end{equation}
to the scaled cumulant generating function
	\begin{align}
	\xi\left(k\right) &= \lim_{T\to \infty} \frac{1}{T} \ln \braket{{e^{Tk G}}},  \label{SCGF}
	\end{align}
	where $\braket{...}$ denotes averaging over the process \eqref{xminos}. According to the DV method, the calculation of $\xi\left(k\right)$ boils down to finding the (minus) ground state energy of a Schr\"{o}dinger-type operator,
	\begin{equation}
	-\psi''/4+V_k\psi=-\xi\psi,\label{eigen}
	\end{equation} 
	with the confining potential given by 
	\begin{eqnarray}\label{selfadj1}
	V_k=-D\cosh 2x+\left(D\sinh2x-r_-\right)^2- 2kD\sinh^2x,\nonumber
	\end{eqnarray}
	see Appendix \ref{dv1} for additional details.
	The solution for the Schr\"{o}dinger problem also provides the biasing potential of the conditioned process \eqref{conm} 
	\begin{equation}
	V_G\left(x_-^c\right)=-\frac{1}{2}\ln \psi_k\left(x_-^c\right),
	\end{equation}
	with the relation $k=k\left(G\right)$ given by the Legendre-Fenchel transform \eqref{GE1}, see e.g. \cite{chetrite_nonequilibrium_2013}. The corresponding steady state distribution for $x_-^c$ , conditioned on a prescribed value $G$ than reads
	\begin{equation}
	P\left(x_-^c\right)=\mathcal N \psi_k^2\left(x_-^c\right),
	\end{equation}
	where $\mathcal N$ is a normalization constant. The value $k=0$ corresponds to the unbiased case where the ground state is given by the unconditional distribution \eqref{xminosdis} $\psi=\sqrt{P_-}$. For increasingly negative $k$ value, the conditional distribution corresponds to increasingly suppressed $G\leq\langle G\rangle$ values. In Appendices \ref{wkb} and \ref{general} we make use of the DV framework to derive some general properties of the population extinction problem.

\section{An alternative derivation using the Fokker-Planck equation}\label{fpsec}
We now show how our hybrid approach also follow from an educated ansatz for the stationary solution $P\left(x_{+},x_{-}\right)$ of the Fokker-Planck equation corresponding to dynamics\eqref{xplos}-\eqref{xminos}.
The stationary Fokker-Planck equation reads
\begin{eqnarray}
0&=&-\partial_{x_{-}}\left[\left(r_{-}-D\sinh2x_{-}\right)P\right]\nonumber\\
&-&\partial_{x_{+}}\left[\left(r_{+}+2D\sinh^{2}x_{-}\right)P\right]
+\frac{1}{4}\nabla^{2}P.\label{eq:fp}
\end{eqnarray}
Our above analysis suggests that the solution should exponentially decrease with $\ln K$ along the sum coordinate
$x_{+}$. However, as opposed to a standard IM ansatz,
the structure along the $x_{-}$ coordinate has an $\mathcal{O}\left(1\right)$
width. The correct ansatz for the quasi steady state distribution then reads 
\begin{equation}
P\simeq h(x_{-})e^{-\ln KS\left(x_{+}/\ln K\right)}.\label{eq:anz}
\end{equation}
Plugging it into Eq.~\eqref{eq:fp}, the results for the extinction exponent $W$ \eqref{extinction}, as well as the conditioned dynamics \eqref{conm} and \eqref{conp} can be reproduced, see Appendix \ref{fp}.
The derivation in Appendix \ref{fp} also provides a way to directly simulate extinction trajectories as in Fig. \ref{comics}, through their time-reversed process.

	\section{Conclusion and broader applicability}\label{conc}
	In this paper we have formulated a hybrid large deviation approach to treat the extinction of coupled populations. Its most striking feature is that the rare extinction state is approached by an ensemble of extinction paths that we have characterized, rather than a single optimal path. We find that extinction probability may be non-monotonic in the migration,
	with the lowest probability of extinction at an intermediate value of $D$. Another surprising behavior we find, is that at small migration $D$ a sink habitat may help to protect the overall population against extinction.
	
	The framework that we have laid out here can be placed within a much broader context. Colored noise, and in particular non-Gaussian noise is often generated by an autonomous process, see e.g. \cite{valenti_noise_2016,barkai_area_2014,hutt_additive_2008,kitada_power-law_2006,walter_first_2020,klosek-dygas_colored_1989,kamenev_how_2008,parker_noise-induced_2011,hakoyama_extinction_2005,evans_stochastic_2013,hening_stochastic_2018,valenti_noise_2016,woillez_nonlocal_2020,woillez_activated_2019,woillez_active_2020,basu_long-time_2019,bouchaud_growth-optimal_2015,yahalom_phase_2019,walter_first_2020,hutt_additive_2008}. A prototypical model of this family features a ``reaction coordinate" $x_+$ driven both by Gaussian noise and a non-Gaussian colored noise $x_-$
	\begin{align}\nonumber
	\dot{x}_{+} & =-U'\left(x_{+}\right)+g\left(x_{-}\right)+\sigma_+\eta_{+},\\
	\dot{x}_{-} & =-V'\left(x_{-}\right)+\sigma_-\eta_{-}.\label{gen}
	\end{align}
	Previous works that examined rare events within the family \eqref{gen} \cite{basu_long-time_2019,abta_amplitude-dependent_2007,hakoyama_extinction_2005,yahalom_phase_2019,kamenev_how_2008,kitada_power-law_2006,klosek-dygas_colored_1989,parker_noise-induced_2011,woillez_nonlocal_2020,woillez_active_2020}, give partial or no analytical analysis, or calculate $W$ for specialized forms of \eqref{gen}, such as linear $V'$ or $g$ \cite{parker_noise-induced_2011,yahalom_phase_2019,kamenev_how_2008,klosek-dygas_colored_1989,woillez_nonlocal_2020,woillez_active_2020}, without addressing the generic case or providing the fluctuating dynamics conditioned on extinction as in Eqs. \eqref{conm}-\eqref{conp}.
	
	We look at rare events of \eqref{gen} in which the reaction coordinate $x_{+}$ reaches a large potential difference $\Delta U/\sigma_+^{2}\!\gg\! 1$.  However, in general, the accompanying $x_-$ trajectories are not characterized by rare $x_-$ state, $\Delta V/\sigma_-^2\!\gg\mkern-21mu / \enskip{} \!1$. Consequently, as in the model examined above, simple rescaling of the $x_-$ coordinates fail to bring the system to a small noise one, and IM fails to account for it during a rare event. 
	
	One exception is the case of linear $V$ and $g$, where $x_-$ is an Ornstein–Uhlenbeck process. Here one can show that the time average $G=\int_0^Tdtg\left(t\right)/T$, over a long times $T$, is a Gaussian variable. As such, it displays the usual IM small noise scaling. As a result, the IM happens to correctly reproduce the large deviation for rare events, however without accounting for the fluctuating  dynamics conditioned on the rare event. Indeed, for this case the DV is reduced to the IM.  This simpler case appeared in several previous works such as \cite{kamenev_how_2008,klosek-dygas_colored_1989}. 
	
	Yet even when $x_-$ is an Ornstein–Uhlenbeck process, once $g$ is nonlinear then the IM is in general inapplicable. For the special case where it is quadratic $g\propto x^2$ then the biased path integral that corresponds to conditioning the process \eqref{gen} on a given time average of $g$ is a Gaussian path integral which can be evaluated exactly. This simplification was used in \cite{parker_noise-induced_2011}, but, without addressing the dynamics.
	Here the corresponding DV problem is easily solved exactly where the eigenfunction problem \eqref{eigen} takes the form of the simple quantum harmonic oscillator, reproducing the results reported in \cite{parker_noise-induced_2011}.
	
	For general non-Gaussian noise (non-linear $V(x),g(x)$), and for addressing the dynamics, one must employ our hybrid DV and IM formalism. It applies when the time scale of the rare $x_{+}$ history is much larger than the relaxation time of $x_-$ in the potential $V$, in which case the accompanying $x_{-}$ histories can be fully characterized by the DV formalism.
	
	For constant $U'\left(x_{+}\right)$, the derivation can be followed directly by substituting $2D\sinh^2x_-\rightarrow g$ and $D\sinh 2x_--r_-\rightarrow V'$. 
	The extension to non-constant $U'(x_+)$ proceeds by dividing the $x_+$ trajectory to small pieces, where $U'$ can be taken as constant, see the Appendix \ref{comp} and also \cite{parker_noise-induced_2011}. In summary, in the broader family \eqref{gen}, multiple paths lead to rare events, and are treated with the same framework.
 	
	\section*{ACKNOWLEDGMENTS}
	G. Bunin acknowledges support by the Israel Science Foundation (ISF) Grant no. 773/18.

	\appendix
	
		\begin{widetext}
		
	\section{Generating extinction trajectories}\label{fokgen}
	Here we show how the solution to the Fokker-Planck equation that corresponds to the process \eqref{lan} can be used to generate trajectories conditioned on extinction. We do so by finding the \textit{time-reversed process}. Here we express it in terms of the $x_{\pm}$ variables \eqref{xplos}-\eqref{xminos} and denote it by $x_+^r,x_-^r$. Initiating that process near the extinction point $x_+=0$ corresponds to the time-reversal of the original process conditioned on extinction. 
	
	The time-reversed Langevin dynamics can be obtained from the steady state distribution
	$P\left(x_{-},x_{+}\right)$ \cite{anderson_reverse-time_1982}: 
	\begin{eqnarray}
	\dot{x}_{-}^{r} & = & -r_-+D\sinh2x_{-}^{r}+\frac{1}{2}\partial_{x_{-}}\ln P+\eta_{-}/\sqrt{2}\label{xminosr}\\
	\dot{x}_{+}^{r} & = & -r_+-2D\sinh^{2}x_{-}^{r}+\frac{1}{2}\partial_{x_{+}}\ln P+\eta_{+}/\sqrt{2},\label{xplosr}
	\end{eqnarray}
	where we omitted terms coming from the regulating terms $h$. 
	To generate trajectories which start at the carrying capacity and are conditioned on extinction, we run the above dynamics starting at extinction $x_+=0$, and than reverse the time of the generated trajectory $x_+^r\left(T-t\right),x_-^r\left(T-t\right)$ where $T$ is the time until stabilization around the carrying capacity $x_+=\ln K$. An instance of such trajectories are presented in Fig.\ref{comics} (B), where we used a numerical solution for the Fokker-Planck equation $P$ with a logistic regulating term. We made sure to add the corresponding terms to the Eqs.\eqref{xminosr}-\eqref{xplosr} as well.
			
\section{IM for the extinction of $x_+$}\label{im}

		In this section we review the derivation of Eqs. \eqref{cond}-\eqref{a} of the main text using the IM.
		
		Given a $x_-$ trajectory, the $x_+$ dynamics \eqref{xplos} is a simple biased diffusion, with a time-dependent drift, whose probability distribution can be found exactly, see Sec. \ref{xp}. However, being interested in the large $K$ limit, we will employ here the IM which can only evaluate the extinction probability up to an exponential pre-factor. The advantage of the IM here, besides its simplicity, is that it can be straightforwardly extended to more involved cases with an $x_+$ dependent forcing. 
		
		We start with the \emph{conditional} probability path measure $P\left[\left\{x_+\left(t\right)\right\}|\left\{x_-\left(t\right)\right\}\right]$ for observing the path history $x_+\left(t\right)$ \emph{given} a $x_-\left(t\right)$ history. It is given, up to pre-exponential factors  $
		P\left[\left\{x_+\left(t\right)\right\}|\left\{x_-\left(t\right)\right\}\right]\propto e^{-S},
		$
		by the conditional path action
		\begin{equation}
		S\left[\left\{x_+\left(t\right)\right\}|\left\{x_-\left(t\right)\right\}\right]=\int_0^T\left(\dot{x}_+-r_+-2D\sinh^2 x_-\right)^2dt'.\label{act}
		\end{equation}
		During an extinction, $x_+$ declines from the large value $\ln K$ to $0$ during time $T$. 
		To prove applicability of the IM we employ the re-scaling 
		$y_{+}=x_{+}/\ln K$ and $\tau=t/\ln K$
		and successfully isolate the large parameter $\ln K$ in front of the conditional
		path measure $S\left[\left\{ y_{+}\left(\tau\right)\right\} |\left\{ x_{-}\left(t\right)\right\} \right]=\ln K\int_0^{T/\ln K}\left(\partial_{\tau}y_{+}-r_{+}-2D\sinh^{2}x_{-}\right)^{2}d\tau$. Importantly, as the optimal extinction duration $T$ scales also with
		$\ln K$ as for the single patch case, then the $\ln K$ dependence
		is pulled of from the integration limit. The large parameter $\ln K$ in front of the conditional path measure makes the IM treatment for the conditional extinction valid here where
		the extinction event is dominated by the minimum action $-\ln P\left[\left\{x_+\left(T\right)=0\right\}|\left\{x_-\left(t\right)\right\}\right]\simeq S^*$, evaluated over the optimal extinction path $x_+$ (with $x_+\left(0\right)=\ln K$ and $x_+\left(t=T\right)=0$) which minimizes the action \eqref{act}, see \cite{freidlin_random_1998,touchette_large_2009}. Standard minimization yields Eqs. \eqref{cond}-\eqref{a} of the main text.
		
			\section{The DV formalism for $x_-$ during extinction}\label{dv1}

		Here we give a brief review of the DV formalism in the context of the present problem. According to the DV method, the scaled cumulant generating function $\xi$ in Eq. \eqref{SCGF} of the main text is the maximal eigenvalue of the operator $\hat{L}^{(k)} \equiv \hat{L} + k g\left(x_-\right)$, which is a tilted version of the Fokker-Planck generator $\hat{L}$
		corresponding to the Langevin equation for the stochastic process \eqref{xminos}. The resulting eigenvalue problem reads
		\begin{equation}
		\frac{1}{4}h''+\left[\left(D\sinh 2x-r_-\right)h\right]'+2kD\sinh^2xh=\xi h,\label{h}
		\end{equation} 
		with the boundary conditions $h\left(x\rightarrow\pm\infty\right)=0$. Here and in the following the prime denotes the derivative with respect to the single argument. A usual protocol here is to make the operator in Eq. \eqref{h} self-adjoint by defining the operator $\mathcal H= e^{-U/2}\hat{L}^{(k)} e^{U/2}$ where in our case $U\left(x\right)=-\int^x4\left(D\sinh 2x-r_-\right)=-4\left(D\cosh^2x-r_-x\right)$. This brings us to an effective Schr\"{o}dinger Eq. \eqref{eigen}
		\begin{equation}
		-\frac{1}{4}\psi''+V_k\psi=-\xi\psi\label{selfadj},
		\end{equation}
		with the confining potential
		\begin{equation}
		V_k=\left(D\sinh2x-r_-\right)^2-D\cosh 2x- 2kD\sinh^2x,\label{pot}
		\end{equation}
		and where $\xi\left(k\right)$ is minus the ground state energy. The two eigenfunctions are related via
		\begin{equation}
		\psi\left(x\right)=e^{2\left(D\cosh^2x-r_-x\right)}h\left(x\right).\label{equi}
		\end{equation}
		An example of the confining potential $V_k$ \eqref{pot}, and its associated ground state $\psi_k$ is presented in Fig. \ref{dvpot}.

		\begin{figure}[]
			\begin{tabular}{ll}
				\includegraphics[width=0.25\textwidth,clip=]{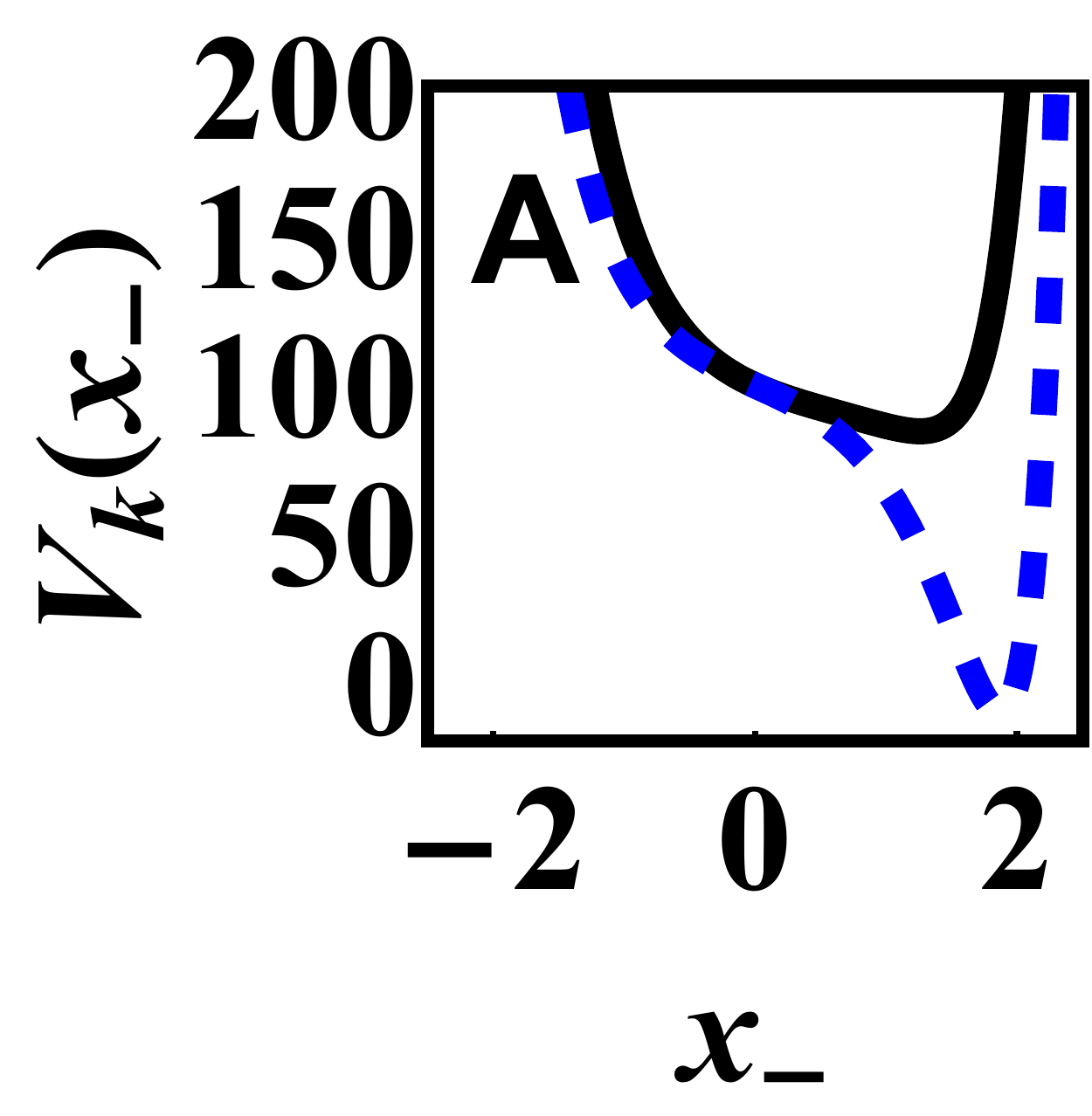}
				\includegraphics[width=0.22\textwidth,clip=]{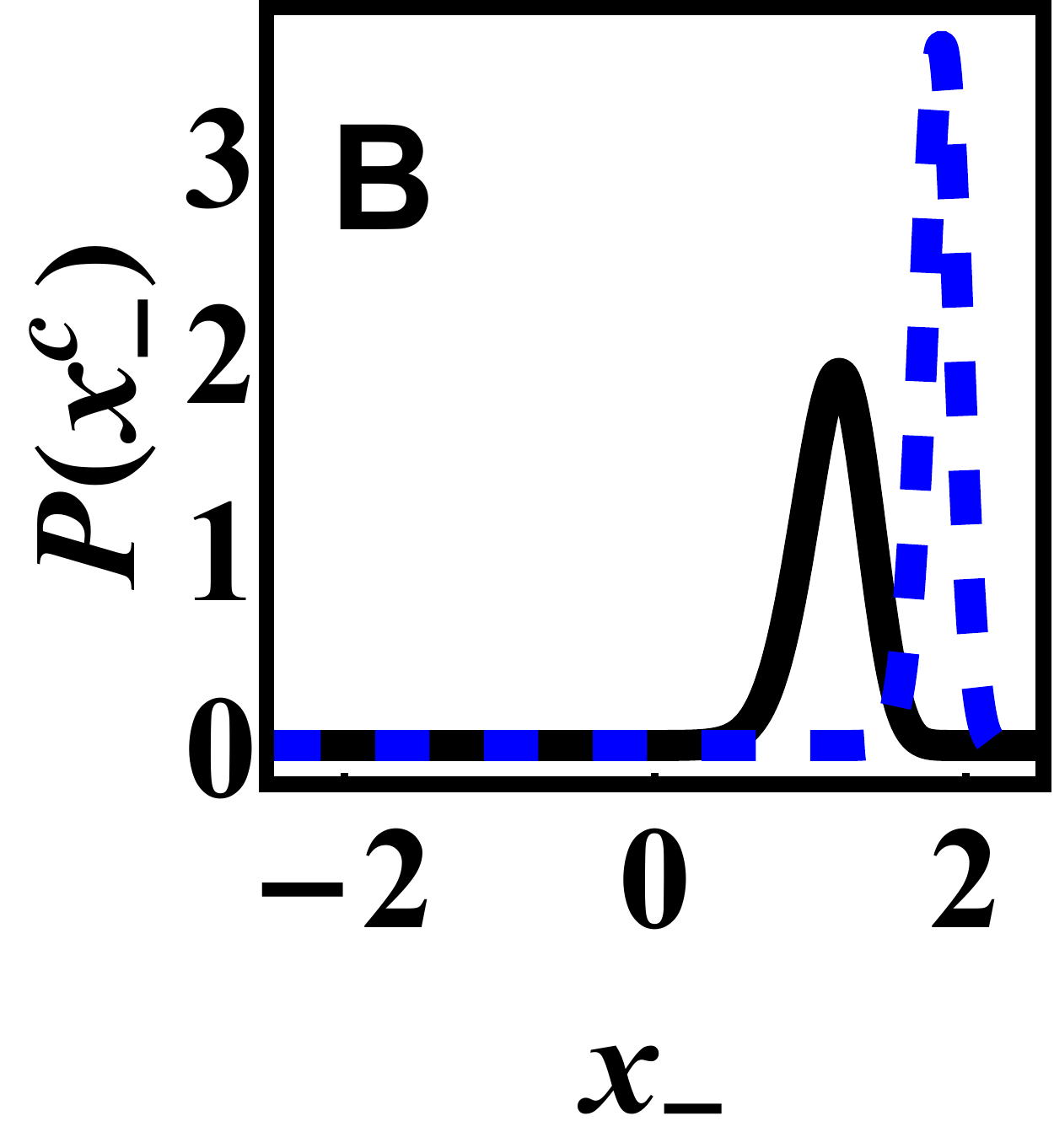}
				\includegraphics[width=0.25\textwidth,clip=]{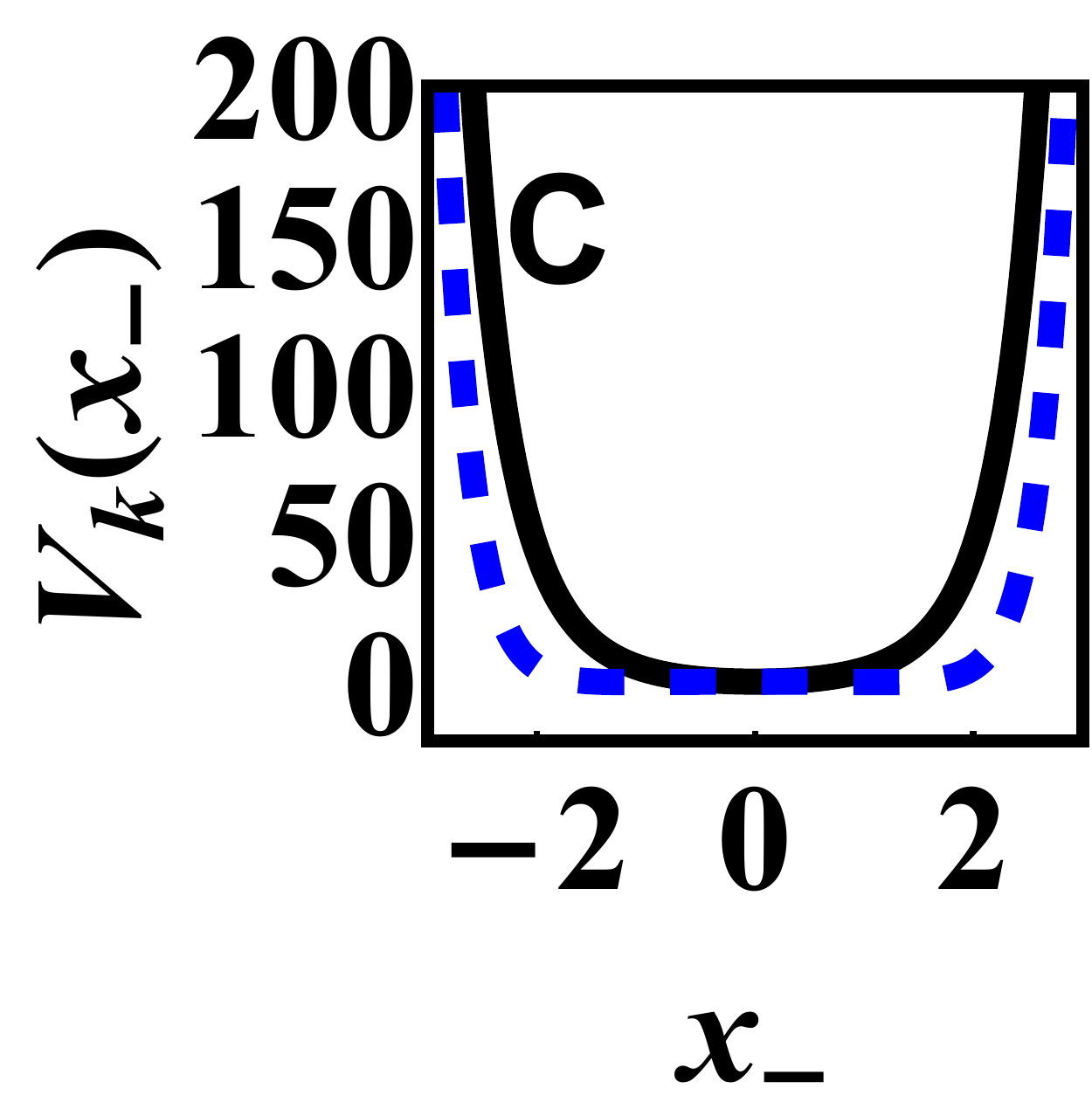}
				\includegraphics[width=0.25\textwidth,clip=]{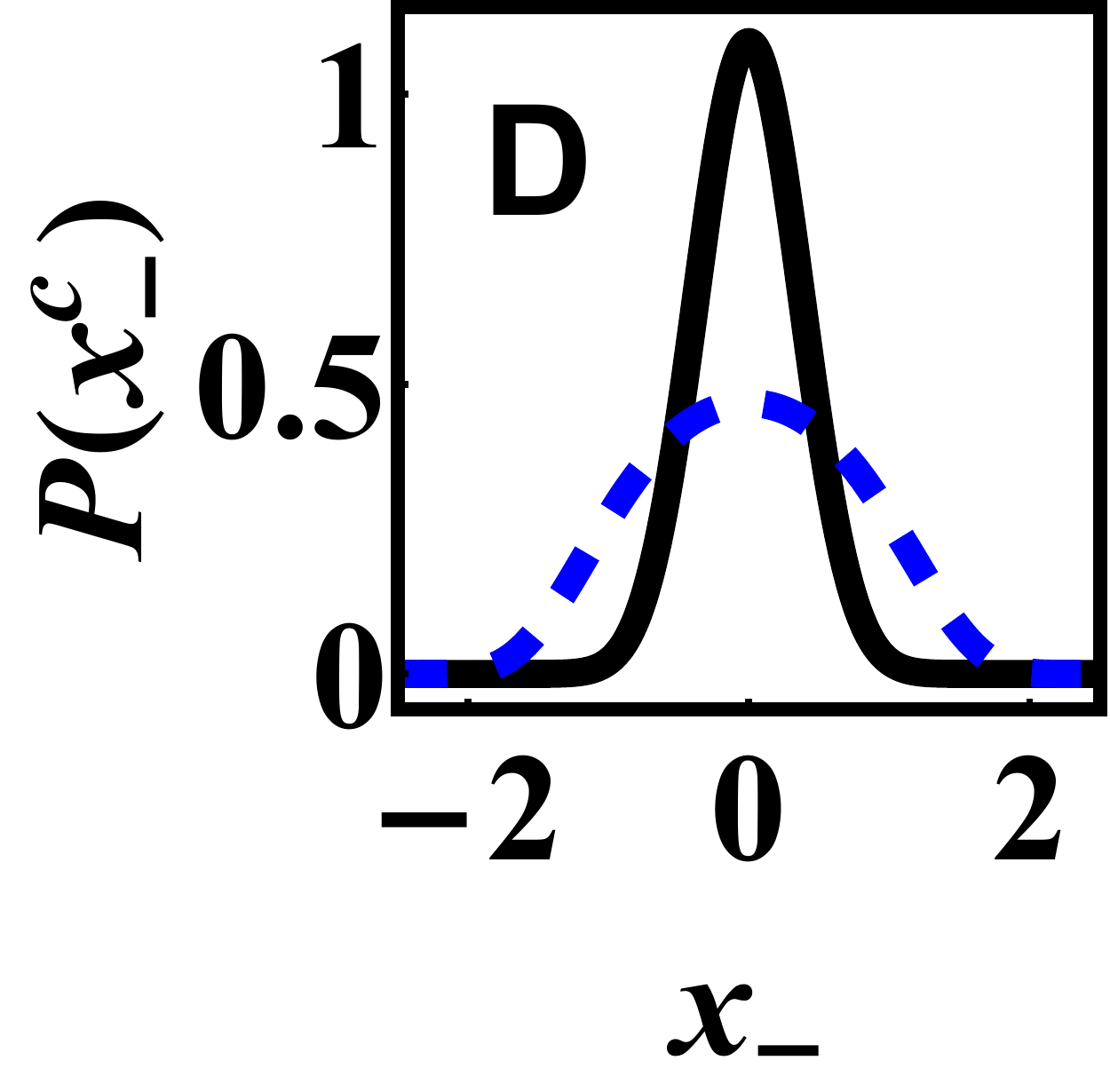}
			\end{tabular}	
			\caption{ The confining potential $V_k$ \eqref{pot} (A),(C) and its associated ground state $\psi_k$ (B),(D) in thick black lines for $k=-15$. The dashed blue line correspond to $k=0$, that is, the unconditional dynamics. In the left panels (A) and (B) $r_-=10$, where the unconditional distribution is peaked away from the origin. For $k<0$, the ground state eigenfunction's peak is advanced toward the origin, thus suppressing the value of $G$ \eqref{a}. In the right panels (C) and (D) $r_-=0$, where the unconditional distribution is symmetric around the origin. The effect of $k<0$ here is to decrease the width of the eigenfunction which again suppresses the value of $G$. In all panels $D=0.1$.  }
			\label{dvpot}		
		\end{figure}

		\section{The hybrid approach reduces to the IM at low noise }\label{wkb}
		At small noise, i.e. $\sigma\ll D,r_{1,2}$ in the dimension-full variables, we have that the dynamics \eqref{lan} take the small noise form in the original variables, without necessitating any rescaling. Thus extinction comes from an IM treatment for both coordinates $x_{\pm}$, and must re-emerge in our hybrid formalism from the DV treatment for the $x_-$ coordinate.
		Indeed, this limit implies, in the rescaled variables that $D,r_{\pm}\gg1$. In such limit, the magnitude of the potential \eqref{pot} of the quantum mechanical problem \eqref{selfadj} diverges, and the ground state energy $-\xi$ is dominated by its minimum $-\xi\simeq\min_{x} \left[\left(D\sinh2x-r_-\right)^2- 2kD\sinh^2x\right]$. Here we retained only leading order terms which scale as $D^2,r_-^2$ (notice that $k\sim D$ in this limit). Performing the Legendre-Fenchel transform \eqref{GE1} one finds
		\begin{equation}
		f\left(G\right)\simeq \left(D\sinh 2x_--r_-\right)^2=\left(\sqrt{G\left(G+2D\right)}-r_-\right)^2\label{fim}
		\end{equation}
		with $x_-=x_-\left(G\right)$ given by the positive solution to
		\begin{equation}
		2D\sinh ^2 x_-=G.\label{gim}
		\end{equation}
		This result is nothing but the IM evaluation for the probability cost of maintaining a long time avarage \eqref{a}. Indeed, the probability path measure for the $x_-$ dynamics \eqref{xminos} is given by  
		$-\ln P\left[\left\{ x_{-}\left(t\right)\right\} \right]\simeq S_- $ with
		\begin{equation}
		S_-\left[ x_{-}\left(t\right) \right]=\int_0^T\left(\dot{x}_{-}+D\sinh2x_{-}-r_{-}\right)^{2}dt'.\label{eq:cond sm}
		\end{equation}
		The optimal path $x_-\left(t\right)$ which minimizes this action \eqref{eq:cond sm} under the constraint of maintaining the time average \eqref{a}, becomes stationary during the long time $T$, and is given by the relation \eqref{gim}. Thus, we find that the DV predictions for the probability cost $-\ln P\left(G\right)\simeq Tf\left(G\right)$ given by Eq.\eqref{fim} and \eqref{gim}, coincides with the IM prediction given by the action $S_-$ \eqref{eq:cond sm} evaluated over the optimal path.
		
		We note here that in the symmetric case $r_-=0$ the IM prediction will always yield $x_-=G=0$ as the optimal value for extinction. Indeed this is the optimal value of $G$ which facilitates extinction of $x_+$ while having the minimal value of the action \eqref{eq:cond sm} $S_-=0$. In the case of noise induced stabilization considered in Sec.\ref{imf}, the optimal $x_+$ trajectory will than follow the deterministic dynamics with zero action \eqref{act} as well, making the combined probability cost vanish.

		\section{Some general properties of the minimization problem \eqref{extinction}}\label{general}
		Here we use details of the DV problem to derive some general properties of the population extinction problem of Sec.\ref{hyb}. 
		
		First, one can conveniently rephrase the minimization problem \eqref{extinction} as a simpler algebraic equation in terms of the cumulant generating function $\xi$ and its argument $k$. Indeed, using the relation $k\left(G\right)=f'\left(G\right)$, and \eqref{GE} one arrives at
		\begin{equation}
		W=-k^*,\label{k}
		\end{equation}
		with $k^*$ the solution to the algebraic equation
		\begin{equation}
		\xi\left(k^*\right)+\frac{k^{*2}}{4}+r_+k^*=0\label{optk}.
		\end{equation}
		
		Second, notice that $W>0$ which implies $f'\left(G^*\right)<0$ which happens for suppressed values of the supplement growth rate $G^*\leq\left<G\right>$. This is in accord with our expectation that during extinction the supplement growth rate will be suppressed in order to facilitate extinction. 
		
		Lastly, one can derive a positive lower bound for the decline rate. From the minimization \eqref{extinction} one concludes that $W$ must always be larger than the value corresponding to the IM probability cost with $G=G^*$, as it also includes the DV probability cost. Thus, one concludes that $W>4\left(r_++G^*\right)$. Using this relation in \eqref{dec} together with $W=-f'\left(G^*\right)$ we find that $r_d>r_++G^*>0$.
		
			\section{The conditioned $x_+^c$ dynamics}\label{xp}
		
		Here we derive the conditioned $x_+^c$ dynamics, Eq. \eqref{conp}.
		
		Starting with Eq. \eqref{xplos} for $x_+$,  define the stochastic variable $y_+=x_+ +r_+\left(T-t\right)+2D\int_t^T\sinh^2x_-\left(t'\right)dt'$, which follows pure Brownian motion
		\begin{equation}
		\dot{y}_+=\frac{1}{\sqrt{2}}\eta_+.
		\end{equation}
		The extinction of $x_+^c$ during time $T$ corresponds to a simple Brownian bridge $y_+\left(t=0\right)=\ln K +\left(r_++G\right)T$, $y\left(t=T\right)=0$, whose conditioned dynamics can be found. e.g., in \cite{rogers_diffusions_1994}
		\begin{equation}
		\dot{y}^c_+=-\frac{y^c_+}{T-t}+\frac{1}{\sqrt{2}}\eta_+.
		\end{equation}
		Notice the deterministic forcing in this equation that enforces the coordinate $y_+^c$ to hit the origin at exactly $t=T$. 
		Transforming coordinates back to $x_+^c$, we find:
		\begin{equation}
		\dot{x}_+^{c}=2D\sinh^2 x_-\left(t\right)-\frac{\int_t^T2D\sinh^2 x_-\left(t'\right)dt'}{T-t}-\frac{x_+}{T-t}+\frac{1}{\sqrt{2}}\eta_+.\label{xp1}
		\end{equation}
		As the accompanying $x_-^c$ trajectories satisfy that the time average of $g$ is equal to $G$ for any macroscopic time interval, than Eq. \eqref{xp1} give way to Eq.\eqref{conp}.

		\section{Typical dynamics in the weak coupling limit $D\rightarrow0^+$}\label{tld}
		
		Here we explain why at small $D$, the typical growth of the total population is dominated by that of patch 1.
		
		At small $D$ the growth of the population at each patch occurs at a different exponential rate $\dot{N}_{i}\simeq \left(r_i+\eta_i\right)N_i$ until the population ratio becomes very large $\mathcal O \left(1/D\right)$. At this point the abundance at the faster patch $1$ is much larger compared to patch 2 and thus migration to patch 2, relative to population size, $D\left(N_2-N_1\right)/N_2$ is non-negligible. The migration to patch 1 is negligible because $N_2$ is small.
		Indeed, the $x_-$ distribution at steady state, Eq. \eqref{xminosdis}, is peaked at $x_-\sim \ln\left(r_-/D\right)$, and the average of migration to patch 1, relative to population size, $\langle D\left(N_2-N_1\right)/N_1\rangle $ with respect to this distribution, vanishes to leading order. The same is true for the variance.
		Thus, the growth of the total population $N_1+N_2$ is dominated by that of patch 1 alone with the average growth rate $r_1$, which coincides with its zero noise value, and so NIS is negligible here.

		\section{Extinction probability in the weak coupling limit $D\rightarrow0^+$}\label{ld}

		Here we present the asymptotic solution of the DV problem \eqref{selfadj} at the small $D$ limit.
	
		The small $D$ limit turns out to be a singular perturbation problem with a sharp transition at the critical value of $k=-2r_-$. Here the confining potential of the Schr\"{o}dinger Eq. \eqref{selfadj}  becomes very wide with the length scale $|\ln D|$ and the solution is given by matched asymptotic expansions \cite{holmes_introduction_2012}. 
		The eigenfunction $\psi$ is characterized by
		a central non vanishing inner boundary layer, 
		flanked by two outer
		boundary tails where it decays to zero. The derivation is rather lengthy and is detailed in the supplementary material. The final results read:\\
		The central boundary layer describing the eigenfunction (for $|x|<|\ln D|$) is given by
		\begin{numcases}
		{\psi_k^2\left(x\right)\simeq}
		\mathcal N_1e^{-4D\left(\cosh^2x-\frac{r_-+\frac{k}{2}}{D}x\right)}, & $-2r_-<k\leq 0$, \label{eigen1}\\
		\frac{\mathcal N_2}{|\ln D|}e^{-4D\cosh^2x}\cos^2\left[\frac{\pi}{|\ln D|}\left(x+\frac{\psi_{DG}\left(-\frac{k}{2}-r_-\right)-\psi_{DG}\left(-\frac{k}{2}+r_-\right)}{4}\right)\right],& $k\leq-2r_-$, \label{eigen2}
		\end{numcases}
		where $\psi_{DG}$ is the DiaGamma function, $\mathcal N_{1,2}$ are $\mathcal O\left(1\right)$ normalization constants and the first line holds away from a vanishing vicinity of $-2r_-$, $k+2r_-=\mathcal O\left(1/\ln D\right)$. These expressions are presented in Fig. \ref{eigenplot} upon substituting $k=-W$, with $W\left(r_1,r_2\right)$ given by \eqref{strong}-\eqref{inter}. $W$ is found using the cumulant generating function together with \eqref{optk}. The derivation of $\xi$ is given in the supplementary material with the final result
		\begin{numcases}
		{\xi\left(k\right)\simeq}
		\frac{k^2}{4}+kr_-, & $-2r_-<k\leq 0$, \label{xl}\\
		-r_-^2,& $k\leq-2r_-$, \label{xh}
		\end{numcases}
		where the first line holds away from a vanishing vicinity of $-2r_-$, $k+2r_-=\mathcal O\left(1/\ln D\right)$, and the second line holds except for $k$ that diverge as $|k|\sim1/D$.
		
		Legendre transforming this expression we arrive at \eqref{lowd}. Accounting for the next order correction to \eqref{xh} (see the supplemental material) one arrives at the leading order correction to $W$ at small $D$ in the extinction source regime $|r_2|\leq r_1$
		\begin{equation}
		W\simeq r_1+r_2+\sqrt{2\left(r_1^2+r_2^2\right)}\left[1+\frac{\pi^2}{4|\ln D|^2\left(r_1^2+r_2^2\right)}\right],\label{lowd1}
		\end{equation} 
	see Fig. \ref{fh}.

		\section{Conditioned dynamics in the weak coupling limit $D\rightarrow0^+$}\label{p}

		In the extinction sink regime $r_2\leq-r_1$, we have that the optimal value of $G$ from Eq.\eqref{opta} is given by $G=-r_+$ which corresponds to $k=-2r_1>-2r_-$. Thus, the corresponding eigenfunction is given by expression \eqref{eigen1}.

		Plugging these results in Eq.\eqref{conm} and \eqref{conp} we find
		\begin{equation}
		\dot{x}_-^c=-r_+-D\sinh 2x_-+\frac{1}{\sqrt{2}}\eta_-,\label{xmc}
		\end{equation}
		and 
		\begin{equation}
		\dot{x}_+^{c}=2D\sinh^2 x_-\left(t\right)+r_+-\frac{x_+}{T-t}+\frac{1}{\sqrt{2}}\eta_+,
		\end{equation}
		from which we obtain
		\begin{eqnarray}
		\dot{x}_1^c&=&-\frac{x_1^c+x_2^c}{2\left(T-t\right)}+D\left(e^{x_2^c-x_1^c}-1\right)+\eta_1,\label{xc1}\\
		\dot{x}_2^c&=&-\frac{x_1^c+x_2^c}{2\left(T-t\right)}+\left(r_1+r_2\right)+D\left(e^{x_1^c-x_2^c}-1\right)+\eta_2.\label{xc2}
		\end{eqnarray}
		Just as for the typical growth, during extinction the abundance in patch 2 is much smaller than in patch 1, and the migration to patch 1 can be neglected. This means we can set the second term in the right hand side of Eq. \eqref{xc1} to zero. Indeed, the $x_-^c$ distribution obtained from Eq. \eqref{xmc} is localized around $x_-^c\sim\ln\left(-r_+/D\right)\gg 1$, and the average, and variance of this term vanishes to leading order.
		Now we also rewrite the Eq. \eqref{xc1} as:
		\begin{eqnarray}
		\dot{x}_1^c=-\frac{x_1^c}{\left(T-t\right)}+r_1\frac{x_-^c}{\ln K\left(1-r_1t/\ln K\right)}+\eta_1\label{x12},
		\end{eqnarray}
		where we also substituted the decline rate $r_d=r_1$.
		As $x_-^c=\mathcal O\left(1\right)$ (do not scale with $\ln K$), than apart from narrow boundary layer in time of width $r_1^{-1}\ll T$ around $t=T$, the $x_-^c$ term can be neglected, and we arrive at the Eq. \eqref{xc12}.
		
		In the same way, we replace the first term on the right-hand side of Eq. \eqref{xc2} by $-x_1^c/\left(T-t\right)$, which can be further approximated by its average $-r_1$. This last statement holds at large $K$. Indeed, the fluctuations of $x_1^c\sim\mathcal O\left(\ln K\right)$ around its average are $\sim\mathcal O\left(1\right)$. As $\left(T-t\right)\sim \mathcal O\left(\ln K\right)$ (except at the end of the process), then fluctuations in $-x_1^c/\left(T-t\right)$ are negligible except for a narrow boundary layer in time around $t=T$. 
		One can in fact provide an explicit proof here. We make use of the exact solution for the $x_1^c$ distribution which follows from Eq.\eqref{xc12}
		\begin{equation}
		P\left(x_1^c,t\right)=\frac{1}{\sqrt{2\pi t\left(1-t/T\right)}}e^{-\frac{\left[x_1^c-r_1\left(T-t\right)\right]^2}{2t\left(1-t/T\right)}},
		\end{equation}
		and so we have that the distribution of $-x_1^c/\left(T-t\right)$ is given by
		\begin{equation}
		P\left[x_1^c/\left(T-t\right)=\tilde{r},t\right]=\frac{1}{\sqrt{2\pi\tilde{\sigma}^2\left(t\right)}}e^{-\frac{\left(\tilde{r}-r_1\right)^2}{2\tilde{\sigma}^2\left(t\right)}},
		\end{equation}
		where
		\begin{equation}
		\tilde{\sigma}^2\left(t\right)=\frac{1}{T}\frac{t/T}{1-t/T}=\frac{r_1}{\ln K}\frac{t/T}{1-t/T}.
		\end{equation}
		Thus, apart from a narrow boundary layer in time around $t=T$ of width $r_1^{-1}\ll T$, the variance scales as $1/\ln K$ and is vanishingly small. Approximating $-\frac{x_1^c+x_2^c}{2\left(T-t\right)}\simeq -r_1$ in the Eq. \eqref{xc2} we arrive at Eq. \eqref{xc22}.

		\section{Extinction in the strong coupling limit $D\rightarrow\infty$}\label{hd}

		Here we present the asymptotic solution of the DV problem \eqref{selfadj} at large $D$.
		
		At large coupling we have that the two patches are infinitely coordinated, and fluctuations in $G$ are significantly suppressed. Consequently we have that the rate function $f$ diverges away from its minimum, $f\simeq \tilde{f}\left(G\right)/\epsilon+\mathcal O \left(1\right)$ where $\epsilon=1/D$ is our small parameter. Correspondingly we have the scaling of the cumulant generating function \eqref{SCGF} $\xi\left(k\right)=\tilde{\xi}\left(\tilde{k}\right)/\epsilon$ with $\tilde{k}=\epsilon k$. The confining potential of the Schr\"{o}dinger Eq. \eqref{selfadj}  becomes vary narrow around its minimum $x_{\text{min}}=\mathcal O \left(\epsilon\right)$ which can be set to zero as it only contributes in the sub-leading order and the eignenfunction $\psi$ is narrowly localized around the origin over a small scale $x\sim1/\sqrt{\epsilon}$.
		Substituting $\tilde{x}=\sqrt{\epsilon}x$, and expanding in powers of $\epsilon$ we have that Eq. \eqref{selfadj} becomes a simple quantum harmonic oscillator
		\begin{equation}
		\frac{1}{4}\partial^2_{\tilde{x}}\psi+\left[1+\left(2\tilde{k}-4\right)\tilde{x}^2+\mathcal O\left(\epsilon\right)\right]\psi=\tilde{\xi}\psi.\label{selfadj6}
		\end{equation}
		The ground state is a Gaussian whose width is parameterized by $\tilde{k}$, which together with the ground state energy reads:
		\begin{equation}
		\psi\left(x\right)\simeq 2\sqrt{\frac{D\alpha}{\pi}}  e^{-2D\alpha x^2}\quad;\quad\tilde{\xi}=1-\alpha+\mathcal{O}\left(\epsilon\right)\quad;\quad \alpha^2=1-\frac{\tilde{k}}{2}.\label{gau3}
		\end{equation}
		Legendre transforming we find the rate function
		\begin{equation}
		f\left(G\right)=\frac{D}{2}\frac{\left(1-G/\langle G\rangle\right)^2}{G/\langle G\rangle}+\mathcal O \left(1\right),\label{ihigh}
		\end{equation}
		where $\langle G\rangle\simeq1/4$ in this limit, see Fig. \ref{fh}(B). As expected, the rate function assigns a diverging cost for $G$ away from the average $\langle G\rangle$. Plugging this result in \eqref{opta} we find to leading order that $G^*=\langle G\rangle$ and
		\begin{equation}
		W=1+4r_+ +\mathcal O\left[\left(\frac{1}{D}\right)
		\right].\label{largedap}
		\end{equation}

		 Sub leading corrections at finite $D$ can be developed in a systematic way. They come from standard quantum mechanical perturbation method applied to the Eq. \eqref{selfadj} where one finds
		\begin{equation}
		W=1+4r_+ - \frac{1}{4D}\left(1 + 2r_+-8r_-^2\right)+\mathcal O\left(\frac{1}{D^2}\right) ,\label{largedap1}
		\end{equation}	
		 This result reveals that $W$ can become a monotonically \textit{decreasing} function of the coupling strength at large coupling, whenever $r_->\sqrt{r_+/4+1/8}$ (and vice verse), see Fig. \ref{fh}(A). 
		 
		 This can be simply understood by examining the average supplement growth rate, and fluctuations around it. On the one hand, increasing the coupling makes fluctuations in the supplement growth less probable. This effect impedes extinction. On the other hand, it also reduces its average which facilitates extinction. The latter effect takes over the first for large growth rate difference $r_-$ where the average supplement growth rate is large. 
		 
		 This effect has an important implication. Indeed, whenever $W$ is monotonically \textit{increasing} at small coupling (as in the extinction source regime \eqref{lowd1}), than the above criterion ensures a non monotonic dependence on $D$ where extinction is least likely at an optimal intermediate coupling strength $0<D^*< \infty $, as evident in Fig. \ref{fh}(A).	A related phenomena was reported in \cite{khasin_minimizing_2012} where the extinction of coupled populations under demographic stochasticity alone was studied. There an optimal migration strength for survival was reported as well. Their findings are quite different than our reporting here, as the optimal value in \cite{khasin_minimizing_2012} vanishes in the large carrying capacity limit. In contrast, we report here an $\mathcal O\left(1\right)$ optimal value for protection against extinction. Indeed, the two models are fundamentally different due to fundamentally different sources for stochasticity.

	\section{Equivalence with the Fokker-Planck approach}\label{fp}
	To derive the results from the Fokker-Planck equation with the ansatz \eqref{eq:anz}, we substitute \eqref{eq:anz} into the stationary Fokker-Planck equation \eqref{eq:fp}, and keep only leading order terms, giving
	\begin{eqnarray}
	A\left(x_-\right)S'^2\left(\frac{x_+}{\ln K}\right)+B\left(x_-\right)S'\left(\frac{x_+}{\ln K}\right)+C\left(x_-\right)=0,\label{heq0}
	\end{eqnarray}
	with
	\begin{eqnarray}
	A=\frac{h}{4}\quad;\quad B=h\left(r_++2D\sinh^2x_-\right)\quad;\quad C=\frac{1}{4}\partial_{x_-}^2h-\partial_{x_-}\left[\left(r_{-}-D\sinh2x_{-}\right)h\right]
	,\label{heq00}
	\end{eqnarray}
and where primes denote derivatives with respect to the argument. Solving the quadratic equation for $S'$ \eqref{heq0} one arrives at a variables separation equality for $S'=F\left(x_-\right)$ and concludes that $F=\text{const}=-W$. Integrating we have $S=-W\left(x_+/\ln K-1\right)$ where we set the integration constant so that the probability \eqref{eq:anz} will be $\mathcal O\left(1\right)$ at $x_+=\ln K$. Thus, the ansatz \eqref{eq:anz} takes the form 
\begin{equation}
P\simeq h(x_{-})e^{W\left(x_+-\ln K\right)},\label{eq:anz2}
\end{equation}
and we see that $W$ corresponds to the extinction exponent \eqref{exprob}. 	
Also, we now have from Eq. \eqref{heq0} that $h\left(x_-\right)$ obeys an eigenvalue-like problem
	\begin{eqnarray}
\frac{1}{4}h''-\left[\left(r_{-}-D\sinh2x_{-}\right)h\right]'-2WD\sinh^{2}x_{-}h
	=-\left(\frac{W^{2}}{4}-r_{+}W\right)h\label{heq}
	\end{eqnarray}
	for the parameter $W$,
	with the boundary conditions $h\left(x_{-}\rightarrow\pm\infty\right)=0$.
	As we show in the appendix \ref{dv1}, under a simple self-adjoining procedure for the operator in the L.H.S of \eqref{heq} this eigenvalue-like problem coincides with the Schr\"{o}dinger Eq. \eqref{eigen}, where the groundstate energy is given by $-\xi=W^2/4-r_+W$. 
	
	This last equality is nothing but the equation for the optimal value of $G^*$ \eqref{opta} rewritten in terms of the dual Legnadre transform parameter \eqref{optk}, upon substituting the relation $W=-k^*$ \eqref{k}. This proves that the Fokker-Planck approach reproduces the predictions of the hybrid approach for the extinction exponent \eqref{extinction}.   
	
	Next, we show how the Fokker-Planck approach can reproduce the dynamics leading to extinction \eqref{conm}-\eqref{conp}. We do so by finding the \textit{time-reversed process} $x_+^r,x_-^r$ corresponding to the dynamics \eqref{xplos} and \eqref{xminos} of the main text. As explained in Sec.\ref{fokgen}, initiating that process near the extinction point corresponds to the time-reversal of the original process conditioned on extinction. 
	
		To make contact with the conditioned dynamics of the hybrid approach \eqref{conm} and \eqref{conp},
	we substitute the ansatz \eqref{eq:anz2} into Eqs.\eqref{xminosr}-\eqref{xplosr} of the reversed process and arrive at:
	\begin{eqnarray}
	\dot{x}_{-}^{r} & = & -r_{-}+D\sinh2x_{-}^{r}+\frac{1}{2}\frac{h'\left(x_{-}^{r}\right)}{h\left(x_{-}^{r}\right)}+\eta_{-}/\sqrt{2}=\frac{1}{2}\partial_{x_-^r}\ln\psi_k\left(x_-^r\right)+\eta_{-}/\sqrt{2}\label{xminosr2}\\
	\dot{x}_{+}^{r} & = & -r_+-2D\sinh^{2}x_{-}^{r}+\frac{W}{2}+\eta_{+}/\sqrt{2}=r_d+G^*-2D\sinh^{2}x_{-}^{r}+\eta_{+}/\sqrt{2}.\label{xplos2}
	\end{eqnarray}
	In the second equality of Eq. \eqref{xminosr2} we used the relation between the eigenfunctions $h$ and $\psi$ \eqref{equi}, and in the second equality of Eq. \eqref{xplos2} we used \eqref{dec}. From the Eq. \eqref{xplos2} we have that the typical time for growth of $x_+^r$ is given by $T=\ln K r_d$.
	  
	  In order to prove equivalence of \eqref{xminosr2}-\eqref{xplos2} with the dynamics \eqref{conm}-\eqref{conp} we need to reverse the time back to its original direction. This is different from the first time reversal above, as the time reversal of $\left(x_+^r,x_-^r\right)$ is conditioned on the final time.	  

	  We start with the dynamics \eqref{xminosr2} which is autonomous and at equilibrium, and therefore remains unchanged under time reversal, and coincides with the conditioned dynamics from our hybrid approach \eqref{conm}.
	  	  
Next, the dynamics \eqref{xplos2} is coupled to the $x_-^r$ dynamics, and for a given instance of $x_-^r$ it is a simple Brownian motion with a time dependent drift. To derive its time-reversed counterpart we define the variable $y^r\left(t\right)=x_+^r+\int_0^t2D\sinh^2x_-^r\left(t'\right)dt'$ which obeys a simple diffusion with a constant drift $\dot{y}^r=r_d+G^* +\eta_+/\sqrt{2}$. Incorporating a regulating term, its dynamics will be equilibrium, and hence its time reversal coincides with the foreword in time dynamics $\dot{y}=r_d+G^*+\eta_+/\sqrt{2}$. The extinction of the foreword dynamics during time $T$ is then given by a Browning bridge $\dot{y}^c=-y^c/\left(T-t\right)+\eta_+/\sqrt{2}$. Going back to the $x_+^c$ dynamics we find
\begin{equation}
\dot{x}_+^c=-\frac{x_+^c}{T-t}-\frac{\int_{t}^T2D\sinh^2x_-^c\left(t'\right)dt'}{T-t}+2D\sinh^2x_-^c\left(t\right)+\frac{\eta_+}{\sqrt{2}},
\end{equation}
where we substituted $x_-^r\left(T-t\right)=x_-^c\left(t\right)$. This is the same equation as \eqref{xp1}, which was shown to be equivalent to Eq. \eqref{conp}. Thus, the conditioned dynamics found within the Fokker-Planck approach coincides with the hybrid formalism predictions, which completes the equivalence between the two approaches.

		\section{Non-constant potential $U'\left(x_+\right)$ in Eq. \eqref{gen}}\label{comp}

		Here we look at rare events for the dynamics 
		\begin{align}\nonumber
		\dot{x}_{+} & =-U'\left(x_{+}\right)+g\left(x_{-}\right)+\sigma_+\eta_{+},\\
		\dot{x}_{-} & =-V'\left(x_{-}\right)+\sigma_-\eta_{-},\label{gens}
		\end{align}
		with non constant $U'$.
		We look at rare events of \eqref{gens} in which the reaction coordinate $x_{+}$ reaches a large potential difference $\Delta U/\sigma_+^{2}\gg 1$.
		For a \textit{given} $x_-\left(t\right)$ trajectory, the probability cost of the $x_+$ histories in this case are given within the IM formalism where one minimizes the \textit{conditional} action of the $x_+$ path probability measure $S_+\left[x_+\left(t\right)|g\left(t\right)\right]=\int dt\left[x_++U'\left(x_{+}\right)-g\left(x_{-}\right)\right]^2/2\sigma_+^2$. However, the accompanying $x_-$ trajectories are not dominated by a single path and IM is inapplicable. 
		
		Instead, if the rare $x_{+}$ history varies over a time scale $\mathcal T$ which is much larger compared to the fast relaxation time $\tau$ of $x_-$ inside the potential $V$, than the accompanying
		$x_{-}$ histories can be found within the DV formalism. 
		Denote the time average $G\left(t\right)=\int_{t-T/2}^{t+T/2}g\left[x_-\left(t'\right)\right]dt'/T$,  over an intermediate time scale $\tau\ll T\ll \mathcal T$ . Than the probability of observing any fluctuations in $G$ decays exponentially with $T$ $-\ln P \simeq T f\left(G\right)$, where the rate function $f$ is given within the DV large deviation formalism. Correspondingly, for any given protocol $G\left(t\right)$, which varies over the slow time scale $\mathcal T$, its probability is evaluated by the time integral $-\ln P \simeq S_{\text{DV}}=\int dt f\left[G\left(t\right)\right]$
		
		Next, as the $x_+$ dynamics is characterized by the same slow time scale, we can safely replace $g\left(x_-\right)$ in the conditional path probability measure by the segmented time average given by $G\left(t\right)$.  Putting everything together, we find that the unconditional probability of observing the large deviation of interest is given by the hybrid minimization problem for the sum of the IM action and DV large deviation function
		$-\ln P \simeq \min_{x_+\left(t\right),G\left(t\right)}\left\{ S_+\left[x_+|G\right]+ S_{\text{DV}}\left(G\right)\right\}$. We see that the usual IM action for the $x_-$ paths is replaced here by the DV large deviation function.
		Most importantly, the DV theory gives access to the histories of the system conditioned on the large deviation of interest.

	\end{widetext}
	
	\bibliography{pre}

\end{document}


\begin{widetext}

		\newpage
		\section*{Supplemental Material to the paper ``Extinctions of coupled populations, and rare-event dynamics under non-Gaussian noise" by T. Agranov and G. Bunin}

Here we present the asymptotic solution of the DV problem at the small $D$ limit. It is given by the groundstate of the eigenvalue problem
	\begin{equation}
-\frac{1}{4}\psi''+V_k\psi=-\xi\psi\label{selfadj},
\end{equation}
with the confining potential
\begin{equation}
V_k=\left(D\sinh2x-r_-\right)^2-D\cosh 2x- 2kD\sinh^2x.\label{pot}
\end{equation}

As reported in the main text, the small $D$ limit turns out to be a singular perturbation problem with a sharp transition at the critical value of $k=-2r_-$. Here the confining potential of the Schr\"{o}dinger Eq. \eqref{selfadj}  becomes very wide with the length scale $|\ln D|$ and the solution is given by matched asymptotic expansions. 
The eigenfunction $\psi$ is characterized by
a central non vanishing inner boundary layer, 
flanked by two outer
boundary tails where it decays to zero. The derivation is rather lengthy. We first cite the final results:\\
The central boundary layer describing the eigenfunction (for $|x|<|\ln D|$) is given by
\begin{numcases}
{\psi_k^2\left(x\right)\simeq}
\mathcal N_1e^{-4D\left(\cosh^2x-\frac{r_-+\frac{k}{2}}{D}x\right)}, & $-2r_-<k\leq 0$, \label{eigen1}\\
\frac{\mathcal N_2}{|\ln D|}e^{-4D\cosh^2x}\cos^2\left[\frac{\pi}{|\ln D|}\left(x+\frac{\psi_{DG}\left(-\frac{k}{2}-r_-\right)-\psi_{DG}\left(-\frac{k}{2}+r_-\right)}{4}\right)\right],& $k\leq-2r_-$, \label{eigen2}
\end{numcases}
where $\psi_{DG}$ is the DiaGamma function, $\mathcal N_{1,2}$ are $\mathcal O\left(1\right)$ normalization constants and the first line holds away from a vanishing vicinity of $-2r_-$, $k+2r_-=\mathcal O\left(1/\ln D\right)$. These expressions are presented in Fig.4 of the main text. The corresponding groundstate energy is given by
\begin{numcases}
{\xi\left(k\right)\simeq}
\frac{k^2}{4}+kr_-, & $-2r_-<k\leq 0$, \label{xl}\\
-r_-^2-\frac{\pi^2}{4|\ln D|^2}+\mathcal O\left(\frac{1}{|\ln D|^3}\right),& $k\leq-2r_-$, \label{xh}
\end{numcases}
where the first line holds away from a vanishing vicinity of $-2r_-$, $k+2r_-=\mathcal O\left(1/\ln D\right)$, and the second line holds except for $k$ that diverge as $|k|\sim1/D$.

We now turn to describe the analytical treatment. 
We solve the eigenvalue problem \eqref{selfadj} separately for $x\geq0$ and $x\leq0$, where at each region we invoke matched asymptotics, and than match the solutions. For each region, the relation $z=2D\sinh^2x$ is invertible, and we can substitute
\begin{equation}
\psi\left(x\right)=e^{-2D\left(\cosh^2x-\frac{r_-}{D}x\right)}\tilde{\psi}_{\pm}\left(2D\sinh^2x\right),
\end{equation}
and arrive at
\begin{equation}
\left(z^2 +2D z\right)\tilde{\psi}_{\pm}''\left(z\right)+\left(z-2z^2+D-4D z\pm 2r_-\sqrt{z^2+2D z}\right)\tilde{\psi}_{\pm}'\left(z\right)+kz\tilde{\psi}_{\pm}\left(z\right)=\xi\tilde{\psi}_{\pm}\left(z\right),\label{exactf2}
\end{equation}
In the following we make expansions in small $D$ and assume that $k$ do not scale with $D$. Letting $k$ diverge $|k|\sim1/D$ will enable to probe the divergence of $f$ at the origin.
\subsubsection{Outer solution}
Here we assume we have a simple expansion of the solution $\tilde{\psi}_\pm\left(z\right)=\tilde{\psi}_0\left(z\right)+D\tilde{\psi}_1\left(z\right)+...$. Substituting this expansion we have
\begin{equation}
z^2\tilde{\psi}_\pm''\left(z\right)+\left(z-2z^2\pm 2r_-\right)\tilde{\psi}_\pm'\left(z\right)+kz\tilde{\psi}_\pm\left(z\right)=\xi\tilde{\psi}_\pm\left(z\right),\label{g1}
\end{equation}
The solution to this equation that does not diverge at infinity, is given by the confluent hypergeometric function $U$
\begin{equation}
\tilde{\psi}_{\pm}\left(z\right)=Az^{\mp r_-+\sqrt{r_-^2+\xi}}U\left[\frac{1}{2}\left(2\sqrt{r_-^2+\xi}\mp 2r_--k\right),1+2\sqrt{r_-^2+\xi},2z\right].
\end{equation}
The other independent solution to the ODE \eqref{g1} is given by the modified Laguerre function which diverges as $e^{2z}$ at infinity and is thus invalid. 
For negative values of $r_-^2+\xi<0$, the outer solution give way to
\begin{equation}
\tilde{\psi}_{\pm}\left(z\right)=A_{\pm}z^{\mp r_-+i\sqrt{-r_-^2-\xi}}U\left[\frac{1}{2}\left(2i\sqrt{-r_-^2-\xi/}\mp 2r_--k\right),1+2i\sqrt{-r_-^2-\xi},2z\right].
\end{equation}
\subsubsection{Inner region}
Now we introduce a boundary layer coordinate $\tilde{z}=Dz$, and expand in the near region 
\begin{equation}
\tilde{\psi}_{\pm}\left(z\right)=v\left(\tilde{z}\right)+D v_1\left(\tilde{z}\right)+...
\end{equation}
Plugging it into \eqref{exactf2} we arrive at
\begin{equation}
\left(\tilde{z}^2+2\tilde{z}\right)v_{\pm}''\left(\tilde{z}\right)+\left(\tilde{z}+1\pm 2r_-\sqrt{\tilde{z}^2+2\tilde{ z}}\right)v_{\pm}'\left(\tilde{z}\right)=\xi v_{\pm}\left(\tilde{z}\right).
\end{equation}
There are two independent solutions to this equation
\begin{eqnarray}
v_{\pm}\left(\tilde{z}\right)&=&\frac{B_{\pm}\left(\sqrt{\frac{\tilde{z}}{2}}+\sqrt{1+\frac{\tilde{z}}{2}}\right)^{2 \sqrt{r_-^2+\xi}}+C_{\pm}\left(\sqrt{\frac{\tilde{z}}{2}}+\sqrt{1+\frac{\tilde{z}}{2}}\right)^{-2 \sqrt{r_-^2+\xi}}}{\left[1+\tilde{z}+\sqrt{\tilde{z}\left(2+\tilde{z}\right)}\right]^{\pm r_-}}
\end{eqnarray}
For negative values of $r_-^2+\xi<0$, the inner solution give way to
\begin{eqnarray}
v_{\pm}\left(\tilde{z}\right)=\frac{D_{\pm}\cos\left[2 \sqrt{-r_-^2-\xi}\arcsinh\left(\sqrt\frac{\tilde{z}}{2}\right)+\varphi_{\pm}\right]}{\left[1+\tilde{z}+\sqrt{\tilde{z}\left(2+\tilde{z}\right)}\right]^{_{\pm}r_-}}.	
\end{eqnarray}
The $\pm$ solutions for $v_{\pm}$ must be matched at $x=0$. As the denominator gives in the original variables
\begin{equation}
\left[1+\tilde{z}+\sqrt{\tilde{z}\left(2+\tilde{z}\right)}\right]^{_{\pm}r_-}=e^{ 2r_-x},
\end{equation}
we must have that
\begin{equation}
B_+=C_-\quad;\quad B_-=C_+\quad;\quad D_+=D_-\quad;\quad\varphi_+=-\varphi_-\label{matchin}
\end{equation}
We are left with matching the outer solutions $\tilde{\psi}_{\pm}$ with the inner ones $v_{\pm}$ by employing a matching condition.
\subsubsection{Matching the inner and outer solutions}
There are two different regimes here:
\subsubsection{$-r_-^2<\xi\leq0$}
In this case we find that the outer solution diverges at the origin. the first two terms are given by:
\begin{eqnarray}
\tilde{\psi}_{\pm}\left(z\rightarrow 0\right)&\simeq&A_{\pm}z^{\mp r_-}\frac{\Gamma\left(-2\sqrt{r_-^2+\xi}\right)}{\Gamma\left[-\frac{1}{2}\left(2\sqrt{r_-^2+\xi/\sigma^2}+k\pm 2r_-\right)\right]}\times\\
&&\left[z^{\sqrt{r_-^2+\xi/\sigma^2}}+z^{-\sqrt{r_-^2+\xi/}}
\frac{2^{-2\sqrt{r_-^2+\xi}}\Gamma\left(2\sqrt{r_-^2+\xi}\right)}{\Gamma\left[\frac{1}{2}\left(2\sqrt{r_-^2+\xi}-k\mp 2r_-\right)\right]}\frac{\Gamma\left[-\frac{1}{2}\left(2\sqrt{r_-^2+\xi}+k\pm 2r_-\right)\right]}{\Gamma\left(-2\sqrt{r_-^2+\xi}\right)}\right],\nonumber
\end{eqnarray}
where $\Gamma$ is the gamma function.
Expanding the inner solution at infinity we find two matching terms:
\begin{eqnarray}
v_{\pm}\left(\tilde{z}\rightarrow\infty\right)&\simeq&
z^{\mp r_-}\left(2D\right)^{\mp r_-+\sqrt{r_-^2+\xi}}
\left[B_{\pm}z^{\sqrt{r_-^2+\xi}}+\frac{D^{2\sqrt{r_-^2+\xi}}}{2^{2\sqrt{r_-^2+\xi}}} C_{\pm}z^{-\sqrt{-r_-^2+\xi}}\right].
\end{eqnarray}
Than the matching condition reads:
\begin{equation}
\frac{\Gamma\left(2\sqrt{r_-^2+\xi}\right)}{\Gamma\left(-2\sqrt{r_-^2+\xi}\right)}\frac{\Gamma\left[-\frac{1}{2}\left(2\sqrt{r_-^2+\xi}+k\pm 2r_-\right)\right]}{\Gamma\left[\frac{1}{2}\left(2\sqrt{r_-^2+\xi}-k\mp 2r_-\right)\right]}=D^{2\sqrt{r_-^2+\xi}}\frac{C_{\pm}}{B_{\pm}}
\end{equation}
Employing the relation \eqref{matchin} we arrive at
\begin{equation}
\left[\frac{\Gamma\left(2\sqrt{r_-^2+\xi}\right)}{\Gamma\left(-2\sqrt{r_-^2+\xi}\right)}\right]^2\frac{\Gamma\left[-\frac{1}{2}\left(2\sqrt{r_-^2+\xi}+k-2r_-\right)\right]}{\Gamma\left[\frac{1}{2}\left(2\sqrt{r_-^2+\xi}-k+2r_-\right)\right]}\frac{\Gamma\left[-\frac{1}{2}\left(2\sqrt{r_-^2+\xi}+k+2r_-\right)\right]}{\Gamma\left[\frac{1}{2}\left(2\sqrt{r_-^2+\xi}-k-2r_-\right)\right]}
=D^{4\sqrt{r_-^2+\xi}}
\end{equation}
Except for vanishing $r_-^2+\xi=\mathcal O\left(1/\ln D\right)$, the leading order in $D$ is achieved when the left hand side vanishes. That is either $2\sqrt{r_-^2+\xi}-k-2r_-=0$ or $2\sqrt{r_-^2+\xi}-k+2r_-=0$. As $k\leq0$, only the first option can hold and we finally have that for $-2r_-<k\leq0$
\begin{equation} \xi\simeq\frac{k^2}{4}+kr_-,\label{xilowd}
\end{equation}
which holds accept for $k$ in small vicinity of $-2r_-$, $k+2r_-=\mathcal O\left(1/\ln D\right)$.

In addition, we have the following matching for the coefficients
\begin{equation}
A_+=B_+\left(\frac{2}{D}\right)^{-r_-+\sqrt{r_-^2+\xi}}\frac{\Gamma\left[-\frac{1}{2}\left(2\sqrt{r_-^2+\xi}+k+ 2r_-\right)\right]}{\Gamma\left(-2\sqrt{r_-^2+\xi}\right)}
\end{equation}
which in light of the relation \eqref{xilowd} is reduced to
\begin{equation}
A_+=B_+\left(\frac{2}{D}\right)^{\frac{k}{2}}.
\end{equation}
We also deduce that
\begin{equation}
C_+=B_+D^{k+2r_-}\frac{\Gamma\left(-k-2r_-\right)\Gamma\left(2r_-\right)}{\Gamma\left(k+2r_-\right)\Gamma\left(-k\right)}
\end{equation}
and from the matching of the $-$ parts we also find
\begin{equation}
A_-=B_-\left(\frac{2}{D}\right)^{2r_-+\frac{k}{2}}\frac{\Gamma\left(-k\right)}{\Gamma\left(-k-2r_-\right)}=B_+2^{2r_-}\left(2D\right)^{\frac{k}{2}}\frac{\Gamma\left(2r_-\right)}{\Gamma\left(k+2r_-\right)}=A_+2^{2r_-}D^{k}\frac{\Gamma\left(2r_-\right)}{\Gamma\left(k+2r_-\right)}.
\end{equation}
Importantly, as $C_+$ is negligible compared to $B_+$, we have that the inner region is approximated by
\begin{equation}\label{psih}
\psi^2\left(x\right)\simeq \mathcal Ne^{-4D\left(\cosh^2x-\frac{r_-+\frac{k}{2}}{D}x\right)},
\end{equation}
This approximation breaks down when $k$ approaches $-2r_-$ as $C_+$ is no longer negligible.

\subsubsection{$\xi\leq-r_-^2$}
In this case we find that the square root $\sqrt{\xi+r_-^2}$ becomes imaginary and the outer solution is expended at the origin by:
\begin{eqnarray}
\tilde{\psi}_{\pm}\left(z\rightarrow 0\right)&\simeq&\frac{A_{\pm}z^{\mp r_-}}{2^{i\sqrt{-\xi-r_-^2}}}\left[\left(2z\right)^{i\sqrt{-r_-^2-\xi}}\alpha_{\pm}^{*}+\left(2z\right)^{-i\sqrt{-r_-^2-\xi}}\alpha_{\pm}\right]+\dots\\
&=&\frac{A_{\pm}z^{\mp r_-}}{2^{i\sqrt{-\xi-r_-^2}}}|\alpha_{\pm}|\cos\left[ \sqrt{-r_-^2-\xi}\ln\left(2z\right)-\arg\alpha_{\pm}+\dots\right]\nonumber
\end{eqnarray}
where
\begin{equation}
\alpha_{\pm}=\frac{\Gamma\left(2i\sqrt{-\xi-r_-^2}\right)}{\Gamma\left[\frac{1}{2}\left(2i\sqrt{-\xi-r_-^2}-k\mp 2r_-\right)\right]}.
\end{equation}
For the inner solution we find
\begin{equation}
v_{\pm}\left(\tilde{z}\rightarrow\infty\right)\simeq D_{\pm}\left(\frac{D}{2}\right)^{\pm r_-}z^{\mp r_-}\cos\left[ \sqrt{-r_-^2-\xi^2}\ln\left(\frac{2z}{D}\right)+\varphi_{\pm}+\dots\right]\label{near2}
\end{equation}
The matching condition than reads 
\begin{eqnarray}\nonumber
&&\arg\alpha_+ +\arg\alpha_-=\\
&&\arg\left[\frac{\Gamma\left(2i\sqrt{-\xi-r_-^2}\right)}{\Gamma\left[\frac{1}{2}\left(2i\sqrt{-\xi-r_-^2}-k- 2r_-\right)\right]}\right]+\arg\left[\frac{\Gamma\left(2i\sqrt{-\xi-r_-^2}\right)}{\Gamma\left[\frac{1}{2}\left(2i\sqrt{-\xi-r_-^2}-k+ 2r_-\right)\right]}\right]=2\sqrt{-r_-^2-\xi}\ln D\label{eigenlow}
\end{eqnarray}
and the phases are given by
\begin{eqnarray}
\varphi_+=-\varphi_-=\frac{\arg\alpha_--\arg\alpha_+}{2}.\label{varphi}
\end{eqnarray}
Notice, that from the matching condition \eqref{eigenlow} we have the scaling
\begin{eqnarray}
\sqrt{-r_-^2-\xi/\sigma^2}\sim\frac{1}{\ln D}.
\end{eqnarray}
Using the above scaling, we can expand the Gamma functions in the left hand side of \eqref{eigenlow}. Carefully collecting terms for the real and imaginary parts, we have the approximate matching condition  
\begin{equation}
\arg\left[-\psi_{DG}\left(-\frac{k}{2}-r_-\right)-2\gamma-\frac{i}{\sqrt{-r_-^2-\xi}}\right]+\arg\left[-\psi_{DG}\left(-\frac{k}{2}+r_-\right)-2\gamma-\frac{i}{\sqrt{-r_-^2-\xi}}\right]=2\sqrt{-r_-^2-\xi}\ln D\label{eigenlow2}
\end{equation}
where $\gamma$ is the Euler–Mascheroni constant, and $\psi_{DG}$ is the DiaGamma function. 
Let us substitute than $\xi=-r_-^2-\frac{\tilde{\xi}}{\left(\ln D\right)^2}$
\begin{equation}
\arg \left[-\psi_{DG}\left(-\frac{k}{2}-r_-\right)-2\gamma+i\frac{\ln D}{\sqrt{\tilde{\xi}}}\right]+\arg \left[-\psi_{DG}\left(-\frac{k}{2}+r_-\right)-2\gamma+i\frac{\ln D}{\sqrt{\tilde{\xi}}}\right]=-2\sqrt{\tilde{\xi}}.\label{x}
\end{equation}
To find $\varphi_+$ \eqref{varphi} we must solve for $\alpha_{+,-}$. To leading order at small $D$ we have that, apart for $k$ close to $-2r_-$, the arguments of $\alpha_{+,-}$ both approach $-\pi/2$ and so  $\varphi_+$ vanishes for $k$ away from a narrow boundary layer around $-2r_-$.
Beyond that boundary layer, the next order expansion reads
$\varphi_+|\ln D|=\frac{\pi}{4}\left[\psi_{DG}\left(-\frac{k}{2}-r_-\right)-\psi_{DG}\left(-\frac{k}{2}+r_-\right)\right]+\mathcal O\left[\left(\frac{1}{\ln D}\right)\right]$. 
All and all, we can conclude that to leading order the central part is given by
\begin{equation}
\psi^2\simeq \frac{\mathcal N}{|\ln D|}e^{-4D\cosh^2x}\cos^2\left[\frac{\pi}{|\ln D|}\left(x+\frac{\psi_{DG}\left(-\frac{k}{2}-r_-\right)-\psi_{DG}\left(-\frac{k}{2}+r_-\right)}{4}\right)\right],
\end{equation}
where $\mathcal N$ is an $\mathcal O\left(1\right)$ normalization constant.

Lastly, from the Eq.\eqref{x} one can also derive the leading order correction to $\xi$ in this regime
\begin{equation}
\xi=-r_-^2-\frac{\pi^2}{4|\ln D|^2}+\mathcal O\left(\frac{1}{|\ln D|^3}\right).
\end{equation} 

	\end{widetext}